\documentclass[aps,prb,twocolumn,superscriptaddress,showpacs,longbibliography]{revtex4-2}
\usepackage[dvipdfmx]{graphicx}
\usepackage{bm}
\usepackage{amsmath}
\usepackage{amssymb}
\usepackage{textcomp}
\usepackage{graphicx}
\usepackage{color}
\usepackage{subfigure}
\usepackage{soul,xcolor}
\usepackage{braket} 

\input{epsf}

\begin{document}
	
\title{Resonant excitation of single and coupled qubits\\ for coherent quantum control and microwave detection}
\author{O.~A.~Ilinskaya}
\affiliation{B.~Verkin Institute for Low Temperature Physics and Engineering of the National Academy of Sciences of Ukraine, Kharkiv 61103, Ukraine}

\author{S.~N.~Shevchenko}
\affiliation{B.~Verkin Institute for Low Temperature Physics and Engineering of the National Academy of Sciences of Ukraine, Kharkiv 61103, Ukraine}

\begin{abstract}
Resonant driving enables coherent control of quantum systems, including single and coupled qubits. From a complementary perspective, transitions of a quantum system can be exploited for the detection of microwave photons. In this work, we theoretically investigate resonant multiphoton excitations in a system of qubits. When the energy of $K$ photons matches the energy splitting of the qubit system, the absorption of these photons leads to collective excitation of the qubits. We focus on the case of two coupled qubits and analyze the quantum dynamics of both excitation and relaxation processes. In the particular case where only a single qubit is relevant and the remaining qubits can be neglected, the dynamics admits an analytical treatment. We examine multiphoton resonances, the Bloch–Siegert shift, and population inversion, phenomena that are central to both coherent quantum control and microwave photon detection.
\end{abstract}
	
\maketitle

\section{Introduction}
Quantum systems with energy levels modulated in time have been intensively studied since the early days of quantum mechanics to the present day~\cite{silveri2017}. A qubit is the simplest quantum system, and superconducting qubits are used in many applications, including quantum computing. A two-qubit system is of particular interest as the simplest system in which entanglement can be realized and processed. One of questions addressed for entangled systems is the suppression of decoherence~\cite{konovalenko2025}. In Ref.~\cite{izmalkov2004} quantitative agreement between experiment and theory  serves as evidence of entanglement between two inductively coupled flux qubits with independently controlled bias fluxes. The creation, destruction, and revival of entanglement in an open system of two periodically strongly driven coupled qubits were studied in Ref.~\cite{gramajo2018}. Strong driving can lead to such an interesting feature as an asymmetric volcano lineshape of an oscillating spin defect in diamond~\cite{antonic2025}.

Superconducting qubits are widely used as microwave photon detectors~\cite{besse2018,dassonneville2020,lescanne2020}. One application of such detectors is the measurement of the state of a superconducting qubit coupled to a qubit-detector~\cite{opremcak2018}. In Ref.~\cite{opremcak2018} the detector is based on a multilevel system --- a flux-biased phase qubit (see also Refs.~\cite{ilinskaya2024,stolyarov2023,stolyarov2025}). Transferring the excited state of the qubit to be measured to the detector leads to a flip of the direction of the magnetic flux threading the detector loop. This flip is due to an interwell tunneling, and it is a ``click" of the detector. If the qubit is in the ground state, then there is no ``click" of the detector. The result of the qubit measurement is therefore stored in a classical state --- the direction of the magnetic flux. We note that the interwell tunneling in an asymmetric double-well potential also occurs in other quantum systems such as single-molecule magnets~\cite{sirenko2025}. Remarkably, in some cases it is sufficient to consider double-well systems in the two-state limit~\cite{savasta2021}.

In the recent proposal~\cite{rettaroli2025} a system of two qubits coupled to the same storage resonator is suggested for use in the detection of microwave photons in order to suppress the dark count rate --- the important characteristics of a photon detector. A system of two coupled SQUIDs was also proposed as a sensitive microwave detector~\cite{rehak2014,neilinger2026}.

In driven quantum systems, multiphoton resonances are of particular interest. If the energy difference between the levels of a driven qubit approximately equals the energy of $K$ photons of the driving signal, one observes $K$-photon resonances in the frequency dependence of the time-averaged occupation probabilities of the qubit levels~\cite{shevchenko2008}. 

Coherent interference can be observed even if the driving frequency is rather small and individual resonances are not distinguishable~\cite{berns2006}. A recent preprint~\cite{huang2025} perturbatively studies multiphoton processes in periodically driven quantum systems, using a superconducting fluxonium qubit as an example. Another preprint~\cite{kohler2026} on multiphoton resonances proves the existence of a hidden symmetry leading to the emergence of exact crossings of quasienergies. Multiphoton resonances have one important application~\cite{gramajo2018_2}: by tuning the system of two driven coupled flux qubits at or near these resonances, it is possible to create or destroy entanglement in the system. Multiphoton excitations can also lead to unwanted ionization of a superconducting qubit~\cite{wang2025,fechant2025}. Rabi-like oscillations can be induced by multiple Landau-Zener-St\"uckelberg-Majorana transitions in a strongly coupled qubit--resonator system, and over a broad parameter range the corresponding dynamics can be quantitatively described by the adiabatic-impulse model, studied for both single- and two-qubit systems~\cite{neilinger2016,shevchenko2014,shevchenko2008_2}. In this way, detailed and accurate control of multiphoton transitions is important for controlling the dynamics of quantum systems.

In this manuscript we study the dynamics of the occupation probabilities for a driven system of two coupled qubits and note that the two-qubit dynamics can be reduced to a single-qubit one. 
Further, we study the response of a two-level system on a resonant multiphoton excitation for the driving amplitude (i) being constant and (ii) being frequency-dependent. The latter is due to coupling of the qubit to a resonator, and the amplitude has a maximum at the resonator frequency, decaying on both sides of it as a Lorentzian function~\cite{oconnell2008}. For a constant amplitude, the multiphoton resonances --- the time-averaged occupation probabilities as functions of the driving frequency --- increase in their height and width with the rise of the amplitude. For an amplitude $A(\omega)$ of Lorentzian form, the scale of the resonances on the abscissa axis is determined by the quality factor, which is inversely proportional to the full width at half maximum of the function $A(\omega)$.

The rest of the manuscript is organized as follows. Section~II describes a driven two-qubit system and Section~III is devoted to a single-qubit case. The last section is devoted to the conclusions. Appendix presents analytical calculations, which describe the multiphoton excitation of a qubit. 

\section{Driven two-qubit system}

\subsection{Bases and Hamiltonian}

\begin{figure}
	\includegraphics[width=\columnwidth]{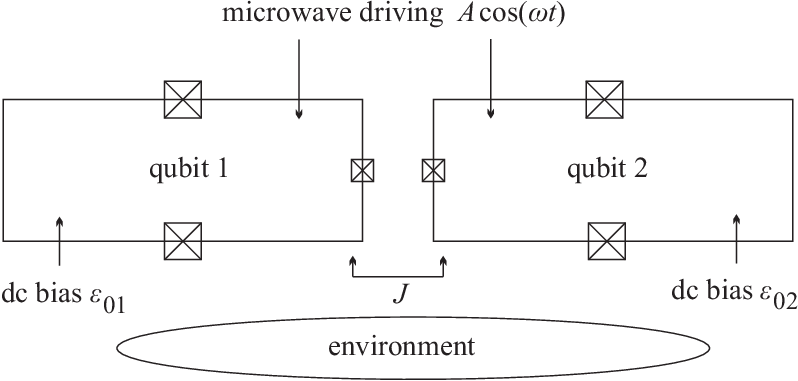}
	\caption{
		Sketch of the model system. Two flux qubits are coupled with the interaction energy $J$. The constant biases are different for the two qubits ($\varepsilon_{01}$ and $\varepsilon_{02}$), while the microwave driving $A\cos(\omega t)$ is the same. The qubits are also coupled to a dissipative environment.  
	}
	\label{Fig:Model}
\end{figure}

We consider the system sketched in Fig.~\ref{Fig:Model}. The Hamiltonian of two coupled superconducting qubits reads~\cite{ploeg2007}
\begin{align}\label{4-level-Hamiltonian}
	&H(t)=-\frac{\Delta_1}{2}\sigma_x\otimes {\hat 1}
	-\frac{\varepsilon_1(t)}{2}\sigma_z\otimes {\hat 1}\nonumber\\
	&\hspace{1.1cm}-\frac{\Delta_2}{2}{\hat 1}\otimes\sigma_x 
	-\frac{\varepsilon_2(t)}{2}{\hat 1}\otimes\sigma_z
	+\frac{J}{2}\sigma_z\otimes\sigma_z\nonumber\\
	&\hspace{0.0cm}=-\frac{1}{2}
	\left(\begin{array}{cccc}
		\varepsilon_+^{-J}(t) & \Delta_2 & \Delta_1 & 0 \\
		\Delta_2 & \varepsilon_-^{+J}(t) & 0 & \Delta_1 \\
		\Delta_1 & 0 & -\varepsilon_-^{-J}(t) & \Delta_2 \\
		0 & \Delta_1 & \Delta_2 & -\varepsilon_+^{+J}(t) \\
	\end{array}\right),
\end{align}
which is written down in the diabatic (flux) basis, formed by vectors
\begin{align}\label{4-level-flux-basis}
	&\ket{\uparrow\uparrow}=\left(\begin{array}{cccc}
		1 \\
		0 \\
		0 \\
		0 \\
	\end{array}\right),\quad
	\ket{\uparrow\downarrow}=\left(\begin{array}{cccc}
		0 \\
		1 \\
		0 \\
		0 \\
	\end{array}\right),\nonumber\\
	&\ket{\downarrow\uparrow}=\left(\begin{array}{cccc}
		0 \\
		0 \\
		1 \\
		0 \\
	\end{array}\right),\quad
	\ket{\downarrow\downarrow}=\left(\begin{array}{cccc}
		0 \\
		0 \\
		0 \\
		1 \\
	\end{array}\right).
\end{align}
Here $\ket{\uparrow\downarrow}=\ket{\uparrow}\otimes\ket{\downarrow}$ etc., and 
\begin{equation}\label{2-level-flux-basis}
	\ket{\uparrow}=\left(\begin{array}{cccc}
		1 \\
		0 \\
	\end{array}\right),\qquad
\ket{\downarrow}=\left(\begin{array}{cccc}
		0 \\
		1 \\
	\end{array}\right)
\end{equation}
are the basis vectors for a single qubit: the vector $\ket{\uparrow}$ corresponds to the supercurrent flowing counterclockwise and the vector $\ket{\downarrow}$ corresponds to the current flowing clockwise.  
In Eq.~(\ref{4-level-Hamiltonian}) $\sigma_x$ and $\sigma_z$ are the Pauli matrices, $\varepsilon_k(t)$ and $\Delta_k$ are the energy biases and tunneling amplitudes for the $k^\text{th}$ qubit, $J$ is the interaction energy, and $\varepsilon_\pm^{\pm J}(t)=\varepsilon_1(t) \pm \varepsilon_2(t) \pm J$ (in this notation the subscript corresponds to the first $\pm$ in the right-hand side, while the superscript corresponds to the second $\pm$). Depending on the sign of $J$, the coupling of the qubits can be either antiferromagnetic-type $(J<0)$ or ferromagnetic-type $(J>0)$. We consider a ferromagnetic coupling below. We take the external excitation into account being identical for both qubits~\cite{gramajo2018,gramajo2018_2}
\begin{equation}
	\varepsilon_k(t)=\varepsilon_{0k}+A\cos(\omega t).
\end{equation}
The angular frequency $\omega$ is in resonance with two eigenstate energy levels out of four ones calculated at the initial moment of time $t=0$, and $A$ is the driving amplitude (in units of energy).

\begin{figure}
	\includegraphics[width=\columnwidth]{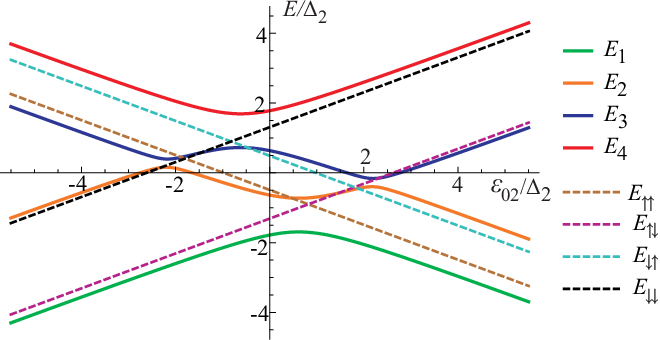}
	\includegraphics[width=\columnwidth]{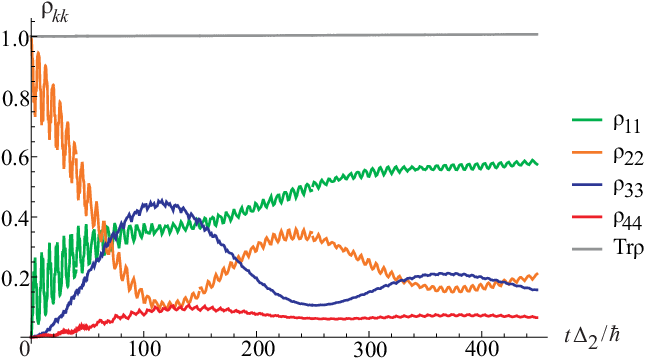}
	\includegraphics[width=\columnwidth]{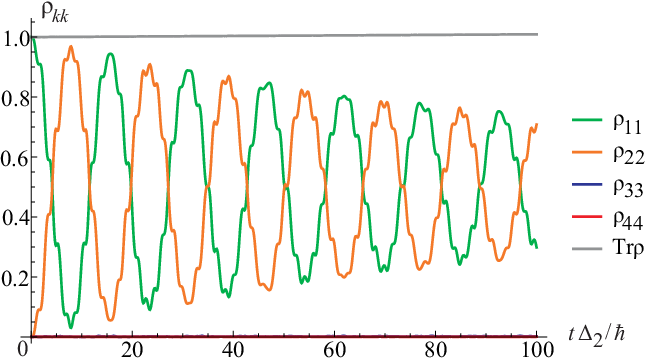}
	\caption{
		Levels and occupations of the microwave driven two-qubit system. 
		Upper panel shows the eigenenergies of the Hamiltonian~(\ref{4-level-Hamiltonian}) (the solid curves) as functions of $\varepsilon_{02}$ at a constant bias $\varepsilon_{01}=2\,\Delta_2$. Other parameters are $\Delta_1=1.5\,\Delta_2$, $J=0.82\,\Delta_2$. The dashed straight lines stand for the energies in the diabatic basis (at $\Delta_1=\Delta_2=0$). 
		Middle and lower panels present the Rabi oscillations calculated in the basis of the eigenfunctions of the stationary Hamiltonian~(\ref{4-level-H0}) with dissipation taken into account. Middle panel corresponds to the frequency of the external signal resonant with the levels 2 and 3, i.e., $\omega=(E_3-E_2)/\hbar$. For lower panel we took $\omega=(E_2-E_1)/\hbar$. Other parameters are $\varepsilon_{02}=4.8\,\Delta_2$, $A=0.2\,\Delta_2$, the dissipation factor $\alpha=0.01$, $k_\text{B}T=\Delta_2$. 
	}
	\label{Fig:TwoQubits}
\end{figure}

The Hamiltonian~(\ref{4-level-Hamiltonian}) can be divided into two parts --- the stationary Hamiltonian 
\begin{align}\label{4-level-H0}
&H_0=-\frac{\Delta_1}{2}\sigma_x\otimes {\hat 1}
-\frac{\varepsilon_{01}}{2}\sigma_z\otimes {\hat 1}\nonumber\\
&\hspace{0.9 cm}-\frac{\Delta_2}{2}{\hat 1}\otimes\sigma_x 
-\frac{\varepsilon_{02}}{2}{\hat 1}\otimes\sigma_z
+\frac{J}{2}\sigma_z\otimes\sigma_z
\end{align}
and the time-dependent part 
\begin{equation}
	V(t)=
	-\frac{A\cos(\omega t)}{2}
	\Big(\sigma_z\otimes {\hat 1}+{\hat 1}\otimes\sigma_z\Big).
\end{equation}
The eigenenergies of the stationary Hamiltonian~(\ref{4-level-H0}) as well as the energies in the diabatic basis are shown in the upper panel of Fig.~\ref{Fig:TwoQubits}.

\subsection{Gorini-Kossakowski-Sudarshan-Lindblad equation}
To describe dynamics, we solve the Gorini-Kossakowski-Sudarshan-Lindblad (GKSL) equation. This fundamental equation is used in various quantum mechanical problems, such as quantum magnetism~\cite{zvyagin2025,zvyagin2025_2}, nanoelectromechanics~\cite{bahrova2022}, quantum optics~\cite{schleich2015}, etc. We switch to the instantaneous (adiabatic) basis, in which the Hamiltonian~(\ref{4-level-Hamiltonian}) is diagonal. The unitary transfer matrix $S(t)$ from the diabatic basis to the instantaneous basis is calculated numerically. The Hamiltonian in the instantaneous basis has the form 
\begin{align}\label{adiabatic-4-level-Hamiltonian}
	&H_\text{i}(t)=
	S^\dag(t)H(t)S(t)\nonumber\\
	&\hspace{0.8cm}=\text{diag}\Big(E_1(t),E_2(t),E_3(t),E_4(t)\Big),
\end{align}
where $E_j(t)$ $(j=1,2,3,4)$ are the eigenenergies of the Hamiltonian~(\ref{4-level-Hamiltonian}). The dissipation is taken into account as in Refs.~\cite{blum2012,temchenko2011}. Then, the GKSL equation in the instantaneous basis reads
\begin{equation}\label{4-level-Lindblad}
	\frac{d\rho_\text{i}(t)}{dt}
	=-\frac{i}{\hbar}
	\left[H_\text{i}(t),\rho_\text{i}(t)\right]
	+
{\cal L}_\gamma\rho,
\end{equation}
where $\rho(t)$ and $\rho_\text{i}(t)=S^\dag(t)\rho(t)S(t)$ are the density operators in the diabatic and instantaneous bases, respectively, and ${\cal L}_\gamma\rho$ is the dissipation superoperator. From this operator equation we obtain the following equation for the matrix elements $\rho_{kk'}(t)\equiv\langle E_{k}(t)|\rho_{\text{i}}(t)|E_{k'}(t)\rangle$ ($\ket{E_k(t)}$ is the eigenvector, $k,k'=1,2,3,4$)
\begin{align}\label{4-level-system-of-equations}
	&\frac{d\rho_{kk'}(t)}{dt}=
	\left(\rho_\text{i}(t)S^{-1}(t)\frac{dS(t)}{dt}+\text{H.c.}\right)_{kk'}\nonumber\\
	&\hspace{1.4cm}-\frac{i}{\hbar}\Big(E_k(t) - E_{k'}(t)\Big)\rho_{kk'}(t)\nonumber\\
	&\hspace{1.4cm}+\delta_{kk'}\sum_{n\neq k'}W_{k'n}\rho_{nn}(t)-\gamma_{kk'}\rho_{kk'}(t).
\end{align}
Here $\delta_{kk'}$ stands for the Kronecker delta, while $\gamma_{mn}$ and $W_{mn}$ are defined as follows
\begin{equation}
\gamma_{mn}=\frac{1}{2}\sum_r (W_{rm}+W_{rn}) - \text{Re}\Gamma_{nnmm} - \text{Re}\Gamma_{mmnn},
\end{equation}
\begin{equation}
W_{mn}=2\text{Re}\Gamma_{nmmn},
\end{equation} 
where
\begin{equation}\label{tensor}
\text{Re}\Gamma_{lmnk}=\frac{1}{8\hbar}\Lambda_{lmnk}J(\omega_{nk})
\left[\coth\left(\frac{\hbar\omega_{nk}}{2 k_\text{B} T}\right)-1\right],
\end{equation}
\begin{equation}
\omega_{nk}=(E_n-E_k)/\hbar,
\end{equation}
\begin{equation}
\Lambda_{lmnk}=\left(\tau_z^{(1)}+\tau_z^{(2)}\right)_{lm}
\left(\tau_z^{(1)}+\tau_z^{(2)}\right)_{nk},
\end{equation}
\begin{equation}
\tau_z^{(1)}=S_\text{st}^{-1}\left(\sigma_z\otimes{\hat 1}\right)S_\text{st},\quad
\tau_z^{(2)}=S_\text{st}^{-1}\left({\hat 1}\otimes\sigma_z\right)S_\text{st},
\end{equation}
$S_\text{st}$ being the transfer matrix from the diabatic basis to the stationary eigenstate basis. In Eq.~(\ref{tensor})  
\begin{equation}
J(\omega)\approx \alpha\hbar\omega
\end{equation}
is the Ohmic spectral density, $T$ is the temperature of the environment, and $k_\text{B}$ is the Boltzmann constant. 
We note that different approaches of taking the dissipation in two-qubit systems into account are described in Refs.~\cite{decordi2017,vadimov2021}.

\subsection{One photon can excite two qubits}
Numerical solution of the GKSL equation allows us to describe dynamics of the two-qubit system. Figure~\ref{Fig:TwoQubits} (middle panel) shows the dynamics for the case when the external signal is resonant with the levels 2 and 3, $\omega=(E_3-E_2)/\hbar$. Here the dynamics of the four-level system cannot be reduced to a two-level one. Similarly, if the signal is in resonance with the diabatic levels $E_{\downarrow\uparrow}$ and $E_{\uparrow\downarrow}$ (see Fig.~\ref{Fig:TwoQubits}, upper panel), both qubits change their states. This was referred to as the process where one photon excites two qubits~\cite{garziano2016}.

\subsection{Two-qubit system can be reduced to a single qubit}
Figure~\ref{Fig:TwoQubits} (lower panel) shows the decaying Rabi oscillations for the external signal being in resonance with the levels 1 and 2, $\omega=(E_2-E_1)/\hbar$. In this case, the four-level system can be reduced to a two-level one. In general, a multilevel system can in some cases be truncated to several levels~\cite{pietikainen2018,reparaz2025}. Similarly, if the microwave signal is resonant with the diabatic levels $E_{\uparrow\uparrow}$ and $E_{\downarrow\uparrow}$ (see Fig.~\ref{Fig:TwoQubits}, upper panel), the state of just one qubit changes.

For a multilevel system without dissipation, the Rabi oscillations are described in Ref.~\cite{likharev}. As shown there, the formula for the occupation probabilities can be obtained without pointing out the explicit form of the stationary Hamiltonian. The perturbation, switched on at the initial moment of time $t=0$, has the form $A\exp(-i\omega t)+\text{h.c.}$. It is assumed that the excitation with the frequency $\omega$ is near-resonant with the two of $N$ levels (with energies $E_n$ and $E_{n'}$), while it is off-resonant with any other pair of levels, i.~e.~
\begin{align}
&|\hbar\omega-(E_n-E_{n'})|\ll\text{min}\Big\{\hbar\omega,\, \nonumber\\ 
&\hspace{0.9cm}|\hbar\omega\pm(E_{n''}-E_{n'})|,\,  |\hbar\omega\pm(E_{n''}-E_n)|\Big\}
\end{align}
$(n''\neq n, n')$.
It is also assumed that at $t=0$ only the level $n'$ is occupied. Then, just the probabilities of occupation of the levels $n$ and $n'$ are nonnegligible and the known Rabi formula for the dynamics of the occupation of the level $n$ is obtained in Ref.~\cite{likharev}. 

So, in the next section we are interested in the case when just one qubit changes its state under the excitation and consider a driven two-level system.

\section{Influence of microwave signal on a single qubit}

\subsection{Hamiltonian and Gorini-Kossakowski-Sudarshan-Lindblad equation}
To describe dynamics of a single qubit, we use in this section the same notations for the Hamiltonian, density operator, eigenenergies, eigenfunctions, etc., as in Section~II. 
We consider the Hamiltonian of a periodically driven qubit, written in the diabatic basis~(\ref{2-level-flux-basis}) as follows
\begin{equation}\label{2-level-Hamiltonian}
	H(t)=
	-\frac{\Delta}{2}\sigma_x
	-\frac{\varepsilon(t)}{2}\sigma_z=
	-\frac{1}{2}\left(\begin{array}{cccc}
		\varepsilon(t) & \Delta \\
		\Delta & -\varepsilon(t) \\
	\end{array}\right),
\end{equation}
where $\Delta$ is the tunneling amplitude and the energy bias is 
\begin{equation}\label{2-level-epsilon}
	\varepsilon(t)=\varepsilon_{0} + A\cos(\omega t).
\end{equation}
The amplitude and angular frequency of the driving signal are denoted by $A$ and $\omega$, respectively, and $\varepsilon_{0}$ is the offset. The Hamiltonian~(\ref{2-level-Hamiltonian}) can also be written as a sum of the stationary Hamiltonian $H_0$  
\begin{equation}\label{2-level-H0}
	H_0=-\frac{\Delta}{2}\sigma_x
	-\frac{\varepsilon_0}{2}\sigma_z=
	-\frac{1}{2}\left(\begin{array}{cccc}
		\varepsilon_0 & \Delta \\
		\Delta & -\varepsilon_0 \\
	\end{array}\right)
\end{equation} 
and the driving Hamiltonian $V(t)$
\begin{equation}\label{2-level-perturbation}
V(t)=-\frac{A}{2}\cos(\omega t)\sigma_z.
\end{equation} 

To write down the GKSL equation, we switch to the instantaneous basis, in which the Hamiltonian~(\ref{2-level-Hamiltonian}) is diagonal. For this, we use the unitary transfer matrix 
\begin{equation}\label{2-level-S}
	\tilde{S}(t)=
	\left(\begin{array}{cccc}
		\gamma_+(t) & \gamma_-(t) \\
		\gamma_-(t) & -\gamma_+(t) \\
	\end{array}\right),
\end{equation}
where 
\begin{equation}
	\gamma_\pm(t)=\frac{1}{\sqrt{2}}
	\sqrt{1\pm\frac{\varepsilon(t)}{\sqrt{\Delta^2+\varepsilon^2(t)}}}.
\end{equation}
Note that $\tilde{S}^\dag(t)=\tilde{S}(t)$.
Then, the new Hamiltonian in the instantaneous basis has the form
\begin{equation}\label{adiabatic-2-level-Hamiltonian}
	H_\text{i}(t)=\tilde{S}(t)H(t)\tilde{S}(t)= 
	\left(\begin{array}{cccc}
		E_-(t) & 0 \\
		0 & E_+(t) \\
	\end{array}\right),
\end{equation}
where $E_\pm(t)=\pm\sqrt{\Delta^2+\varepsilon^2(t)}/2$ are the eigenvalues of the Hamiltonian~(\ref{2-level-Hamiltonian}).

The GKSL equation in the instantaneous basis reads
\begin{align}\label{2-level-Lindblad}
	&\frac{d\rho_\text{i}(t)}{dt}
	=-\frac{i}{\hbar}\left[H_\text{i}(t),\rho_\text{i}(t)\right]\nonumber\\
	&\hspace{0.5cm}+\frac{\Gamma}{2}\Big(2\rho_{++}(t)|E_{-}(t)\rangle\langle E_{-}(t)|
	-|E_{+}(t)\rangle\langle E_{+}(t)|\rho_\text{i}(t)\nonumber\\
	&\hspace{1.4cm}-\rho_\text{i}(t)|E_{+}(t)\rangle\langle E_{+}(t)|\Big),
\end{align}
where $\rho(t)$ and $\rho_\text{i}(t)=\tilde{S}(t)\rho(t)\tilde{S}(t)$ are the density operators in the diabatic and instantaneous bases, respectively, $|E_\pm(t)\rangle$ is the eigenvector of the Hamiltonian~(\ref{2-level-Hamiltonian}), corresponding to the eigenvalue $E_\pm(t)$, 
$\rho_{++}(t)=\langle E_{+}(t)|\rho_\text{i}(t)|E_{+}(t)\rangle$, 
and $\Gamma$ is the relaxation rate. From the operator equation~(\ref{2-level-Lindblad}) we obtain the following differential equations for the matrix elements 
$\rho_{kk'}(t)\equiv\langle E_{k}(t)|\rho_\text{i}(t)|E_{k'}(t)\rangle$ ($k,k'=\{+,-\}$)
\begin{align}\label{2-level-system-of-equations}
	&\frac{d\rho_{kk'}(t)}{dt}
	=-\frac{i}{\hbar}\Big(E_{k}(t)-E_{k'}(t)\Big)\rho_{kk'}(t)\nonumber\\
	&\hspace{0.3cm}-\left(\tilde{S}(t)\frac{d\tilde{S}(t)}{dt}\rho_\text{i}(t)
    +\text{H.c.}\right)_{kk'}\nonumber\\
	&\hspace{0.3cm}
	+\frac{\Gamma}{2}\Big(2\rho_{++}(t)\delta_{k-}\delta_{k'-}
	-\delta_{k+}\rho_{+k'}(t)-\delta_{k'+}\rho_{k+}(t)\Big).
\end{align}

We assume the initial condition in the diabatic basis~(\ref{2-level-flux-basis}), namely, $\rho_{\uparrow\uparrow}(0)=1$ and other matrix elements are equal to zero. Then, the initial condition in the instantaneous basis is as follows: $\rho_{--}(0)=\gamma_+^2(0)$, $\rho_{+-}(0)=\gamma_+(0)\gamma_-(0)$. When one is interested in a stationary solution, as below, the initial condition does not influence the result.

\subsection{Stationary solution and multiphoton resonances}

We solve the system of differential equations~(\ref{2-level-system-of-equations}) numerically. Using the transfer matrix~(\ref{2-level-S}), we express the probability $\rho_{\downarrow\downarrow}(t)$ in the diabatic basis [note that in this section $\rho_{\downarrow\downarrow}(t)=\bra{\downarrow}\rho(t)\ket{\downarrow}$ and it is not related to the two-qubit basis vector $\ket{\downarrow\downarrow}$ in Eq.~(\ref{4-level-flux-basis})] in terms of the matrix elements of the density operator in the instantaneous basis: 
\begin{align}\label{probability-first-formula}
&\rho_{\downarrow\downarrow}(t)=\frac{1}{2}\Big\{1-[\gamma^2_+(t)-\gamma^2_-(t)]
[\rho_{--}(t)-\rho_{++}(t)]\nonumber\\
&\hspace{2.1cm}-4\gamma_+(t)\gamma_-(t)\text{Re}\rho_{+-}(t)\Big\}.
\end{align}
Here, the matrix element $\rho_{\downarrow\downarrow}(t)$ is also the function of all the parameters of the system ($\varepsilon_0$, $\Delta$, $A$, $\omega$), which is not mentioned explicitly. In what follows, we will be interested in the dependence of the matrix elements on the (angular) frequency $\omega$. In Eq.~(\ref{formula-for-averaging}) below we mention this dependence of the level occupation probability explicitly, $\rho_{\downarrow\downarrow}(t)\equiv\rho_{\downarrow\downarrow}(t,\omega)$. 
We average this probability over time, i.e.,
\begin{equation}\label{formula-for-averaging}
	{\overline\rho}_{\downarrow\downarrow}(\omega)=\frac{1}{T_\text{max}-T_\text{min}}
	\int_{T_\text{min}}^{T_\text{max}} dt\,\rho_{\downarrow\downarrow}(t,\omega),
\end{equation}
where $T_\text{min}\sim \Gamma^{-1}$, $T_\text{max}-T_\text{min}\gg\Omega_\text{R}^{-1}$, $\Omega_\text{R}$ being the Rabi frequency. We note that ${\overline\rho}_{\uparrow\uparrow}(\omega)=1-{\overline\rho}_{\downarrow\downarrow}(\omega)$.

Also, a periodically driven single-qubit system can be described analytically in some limiting cases. Following Ref.~\cite{shevchenko_book}, we consider the strong excitation limit (see also Ref.~\cite{oliver2005}, where the Markovian equations are solved in this limit) and suppose that the following conditions are fulfilled:
\begin{equation}\label{conditions}
K \hbar\omega\approx\delta E, \qquad \frac{\Delta}{\sqrt{A\cdot \hbar\omega}}\ll 1.
\end{equation}
The former condition means that the energy of $K$ photons is close to the qubit energy $\delta E=\sqrt{\Delta^2+\varepsilon_0^2}$ (the multiphoton excitation) and allows applying the rotating-wave approximation (RWA). The latter condition is needed~(\cite{ivakhnenko2023}, p.~26) to justify the formula~(\ref{BS-shift}) below for the Bloch-Siegert shift. 

The stationary solution of the Bloch equation gives (see Appendix and Ref.~\cite{shevchenko_book})
\begin{align}\label{rho++}
	\overline{\rho}_{++}(\omega)=\frac{\Delta^2 J_K^2(A/\hbar\omega)/2}
	{\Delta^2 J_K^2(A/\hbar\omega)+(K \hbar\omega-\varepsilon_0)^2 T_2/T_1+\hbar^2/T_1T_2},
\end{align}
where $J_K(A/\hbar\omega)$ denotes the Bessel function of the first kind. 
Here, we are interested in the time-averaged level occupation probabilities in the physical (diabatic) basis~(\ref{2-level-flux-basis}). To obtain these probabilities, see Eq.~(\ref{averaged-probability}) below, we solve in Appendix the Bloch equations and obtain the stationary solution~(\ref{X-stationary}--\ref{Z-stationary}). 


Taking into account Eq.~(\ref{variables-X-Y-Z}), we rewrite Eq.~(\ref{probability-first-formula}) in terms of functions $X$, $Y$, and $Z$, which parameterize the density matrix and are defined in Eq.~(\ref{variables-X-Y-Z}), as follows
\begin{align}\label{probability-second-formula}
&\rho_{\downarrow\downarrow}(t)=\frac{1}{2}\Big\{1-[\gamma^2_+(t)-\gamma^2_-(t)]
Z\nonumber\\
&\hspace{1cm}-2\gamma_+(t)\gamma_-(t)[\cos(K\omega t)X + \sin(K\omega t)Y]\Big\}.
\end{align}
Next, we need to calculate the time-averaged probability ${\overline\rho}_{\downarrow\downarrow}(\omega)$, see Eq.~(\ref{formula-for-averaging}). For this, we substitute the stationary values~(\ref{X-stationary}--\ref{Z-stationary}) in Eq.~(\ref{probability-second-formula}) and note that $\rho_{\downarrow\downarrow}(t)$ is $2\pi/\omega$-periodic. Then, the integral over the large time interval in Eq.~(\ref{formula-for-averaging}) is reduced to the integral over $2\pi/\omega$. Finally we obtain
\begin{equation}\label{averaged-probability}
{\overline\rho}_{\downarrow\downarrow}(\omega)
=\frac{1}{2}\Big(1-I_1\,Z-I_2\,X\Big),
\end{equation}
where $X$ and $Z$ are given by Eqs.~(\ref{X-stationary}, \ref{Z-stationary}) and
\begin{align}
&I_1=\frac{1}{2\pi}\int_{0}^{2\pi}
\frac{\varepsilon_0+A\cos \tau}{\sqrt{\Delta^2+(\varepsilon_0+A\cos \tau)^2}}d\tau,\\
&I_2=\frac{1}{2\pi}\int_{0}^{2\pi}
\frac{\Delta\cos(K \tau)}{\sqrt{\Delta^2+(\varepsilon_0+A\cos \tau)^2}}d\tau.
\end{align}

\begin{figure}[t]	
	\includegraphics[width=\columnwidth]{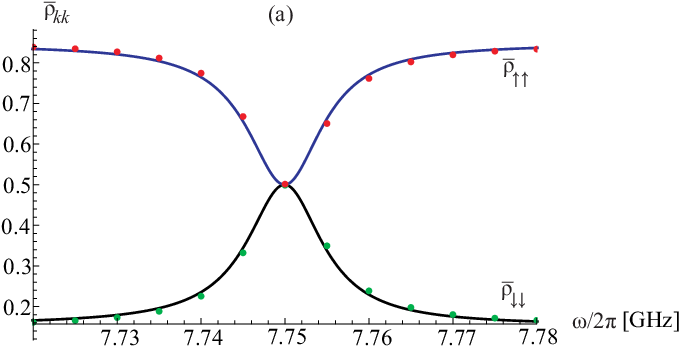}
	\hspace{0.5cm}
	\includegraphics[width=\columnwidth]{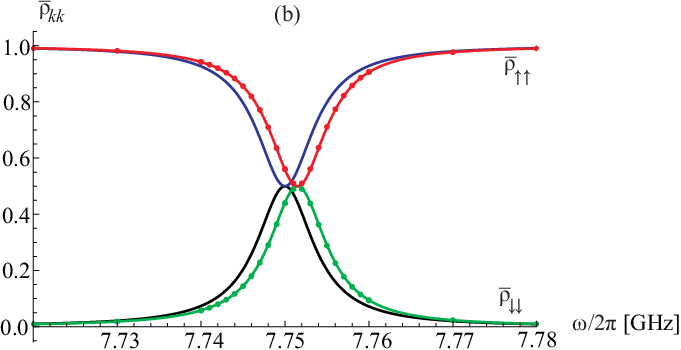}
	\caption{
		The averaged over time occupations (in the diabatic basis) as functions of the driving frequency $\omega$. The circles correspond to the numerical solution of Eq.~(\ref{2-level-system-of-equations}), while the black and blue curves stand for ${\overline\rho}_{\downarrow\downarrow}$ and ${\overline\rho}_{\uparrow\uparrow}$, respectively, calculated analytically by using Eq.~(\ref{averaged-probability}).
		For panel~(a) the parameters are chosen so that the RWA works rather well, namely, the offset $\varepsilon_0=62\,\text{GHz}\cdot h$, the driving amplitude $A=70\,\text{GHz}\cdot h$, the tunneling amplitude $\Delta=0.2\,\text{GHz}\cdot h$, the photon number $K=8$, and the relaxation rate $\Gamma=5\cdot 10^{-3}\,\text{GHz}\cdot h$. 
		In panel~(b) the green and red curves are obtained from the analytical solution~(\ref{averaged-probability}) by shifting the abscissa by $s(\tilde{n})$ [the Bloch-Siegert shift, see the discussion below Eq.~(\ref{least-squares-method})], and this shifted analytical solution coincides with the numerical solution (circles). The parameters used are the following: $\varepsilon_0=62\,\text{GHz}\cdot h$, $A=45\,\text{GHz}\cdot h$, $\Delta=1\,\text{GHz}\cdot h$, $K=8$, and $\Gamma=5\cdot 10^{-3}\,\text{GHz}\cdot h$. }
	\label{Fig:Comparison}
\end{figure}

\begin{figure}
	\includegraphics[width=\columnwidth]{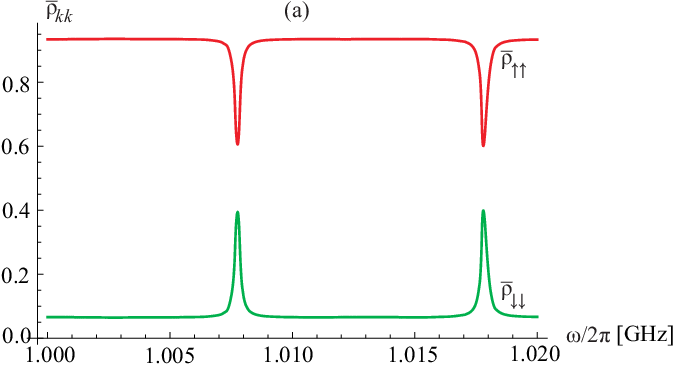}
	
	\vspace{0.2cm}
	\includegraphics[width=\columnwidth]{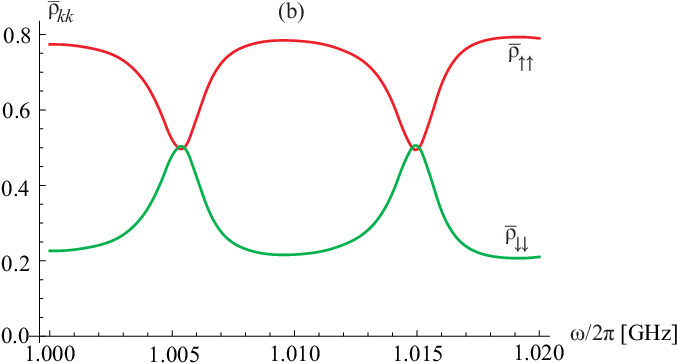}
	
	\vspace{0.2cm}
	\includegraphics[width=\columnwidth]{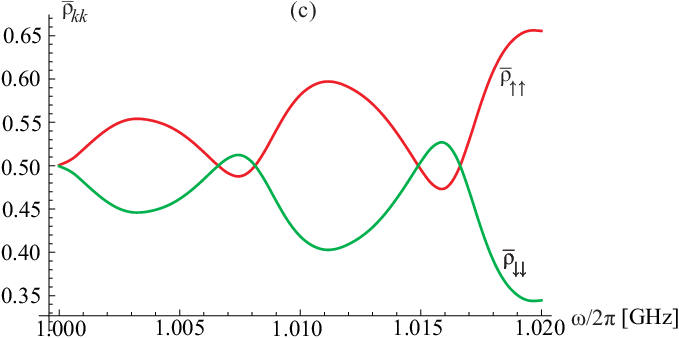}
		
	\vspace{0.2cm}
	\includegraphics[width=\columnwidth]{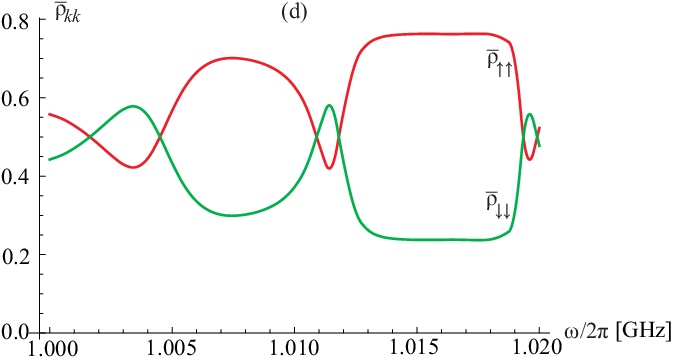}
	\caption{ The averaged over time qubit occupation probabilities as functions of the driving frequency for different values of the driving amplitude. The green and red colors correspond to ${\overline\rho}_{\downarrow\downarrow}$ and ${\overline\rho}_{\uparrow\uparrow}$. Top to bottom:  $A=100\,\text{GHz}\cdot h$, $A=120\,\text{GHz}\cdot h$, $A=150\,\text{GHz}\cdot h$, $A=200\,\text{GHz}\cdot h$. Other parameters are the following: the offset $\varepsilon_0=100\,\text{GHz}\cdot h$ (which corresponds to the photon number $K\approx 100$), the tunneling amplitude $\Delta=10\,\text{GHz}\cdot h$, and the relaxation rate $\Gamma=0.01\,\text{GHz}\cdot h$. } \label{Fig:averages}
\end{figure}

We compare the averaged occupations~(\ref{formula-for-averaging}), obtained by solving the system of equations~(\ref{2-level-system-of-equations}) for the density matrix elements, with the analytical result~(\ref{averaged-probability}). The rotating-wave approximation works rather well when $\Delta\ll\hbar\omega$ (Ref.~\cite{ivakhnenko2023}, p.~28), which is illustrated by Fig.~\ref{Fig:Comparison}(a). We take realistic parameters in Figs.~\ref{Fig:Comparison}--\ref{Fig:highQ} of the same order as in Refs.~\cite{shevchenko2014,neilinger2016}.

\subsection{Multiphoton Bloch-Siegert shift}
When the condition $\Delta\ll\hbar\omega$ is not valid, while conditions~(\ref{conditions}) are still fulfilled, there is a shift between the position of the peak in analytical (obtained by RWA) and numerical curves. This is the Bloch-Siegert shift $\delta_K$ given in the case of a $K$-photon resonance by the formula
\begin{equation}\label{BS-shift}
\delta_K=\frac{\Delta^2}{2 h K}
\sum_{l\neq -K}\frac{J^2_l(A/\hbar\omega)}{\varepsilon_0+l \hbar\omega}.
\end{equation}
Here $h=2\pi\hbar$, $K$ has the same meaning as in Eq.~(\ref{conditions}), and $\delta_K$ has units of frequency. We note that Eq.~(\ref{BS-shift}) differs from Eq.~(75) in Ref.~\cite{ivakhnenko2023} for the Bloch-Siegert shift by the factor $1/K$. That is because these two equations represent shifts for different variables --- $\omega/2\pi$ and $\varepsilon_0$, respectively. 

Figure~\ref{Fig:Comparison}(b) illustrates the existence of the shift between the numerical~(\ref{formula-for-averaging}) and analytical~(\ref{averaged-probability}) results. We calculate this shift by using the least squares method, compare this result with the result obtained by using Eq.~(\ref{BS-shift}), and find a good quantitative agreement. 
Specifically, we seek for the minimum of the function
\begin{equation}\label{least-squares-method}
F(n)=\frac{1}{M+1}\sum_{j=0}^M
\Big\{{\overline\rho}^{\text{(a)}}_{\downarrow\downarrow}(\omega_j-s(n))-
{\overline\rho}^{\text{(n)}}_{\downarrow\downarrow}(\omega_j)\Big\}^2
\end{equation} 
on the grid $n=1,2,...,N$. Here the first term in the curly brackets denotes the analytical solution~(\ref{averaged-probability}) with the abscissa $\omega_j$ shifted to the right by $s(n)=\zeta n$, $\zeta$ being a sufficiently small parameter, and $N$ is chosen so that $\zeta N$ exceeds the difference between the maxima points of ${\overline\rho}^{\text{(n)}}_{\downarrow\downarrow}(\omega_j)$ and ${\overline\rho}^{\text{(a)}}_{\downarrow\downarrow}(\omega_j)$. The second term denotes the numerical solution~(\ref{formula-for-averaging}). (The superscripts are added for clarity.) The values $\omega_j$ are as follows $\omega_j=2\pi\times[7.72+(7.78-7.72)j/(M+1)]\,\text{GHz}$ [see Fig.~\ref{Fig:Comparison}(b)], $M$ being sufficiently large. 
We find $\tilde{n}$ such that $F(\tilde{n})=\text{min}F(n)$. In Fig.~\ref{Fig:Comparison}(b) we show the numerical result~(\ref{formula-for-averaging}) and the result from Eq.~(\ref{averaged-probability}) with the abscissa shifted by $s(\tilde{n})$ and find a good quantitative agreement between $s(\tilde{n})$ calculated by the least squares method and $\delta_K$.

\subsection{Population inversion}

A particularly interesting case is when a population inversion can be realized, i.~e.~when $\overline{\rho}_{\downarrow\downarrow}>\overline{\rho}_{\uparrow\uparrow}$. (Of course, in the instantaneous basis, the population inversion cannot be realized and  $\overline{\rho}_{++}\leq\overline{\rho}_{--}$.)
To consider this case, we find the numerical solution of the GKSL equation~(\ref{2-level-system-of-equations}) for the set of parameters where the second condition in Eq.~(\ref{conditions}) breaks down and the analytical description cannot be applied. We take different values of the driving amplitude $A$ (see Fig.~\ref{Fig:averages}) and show that in the physical basis, the ``inverse" population is possible for sufficiently large $A$, that is, ${\overline\rho}_{\downarrow\downarrow} > 1/2$ [see Fig.~\ref{Fig:averages}(c, d)]. We now comment on the positions of the resonances, which can be clearly distinguished in Fig.~\ref{Fig:averages}(a, b), and the distance between them. Setting the second term in the denominator of Eq.~(\ref{rho++}) to zero, one defines the positions of the resonances as follows
\begin{equation}\label{positions-of-resonances}
	\hbar\omega_K\approx\delta E/K,
\end{equation}
where the tunneling amplitude $\Delta$ is small and $\delta E\approx \varepsilon_0$.
Then, the distance between the adjacent resonances for large $K$ is
\begin{equation}\label{distance-between-resonances}
	\hbar\omega_K-\hbar\omega_{K-1}\approx\delta E/K^2.
\end{equation}
Position of the resonances in Fig.~\ref{Fig:averages}(a, b) and the distance between them are in agreement with Eqs.~(\ref{positions-of-resonances}--\ref{distance-between-resonances}). 
We note also that the width of the multiphoton resonances grows with an increase of the driving amplitude. If one considers the qubit as a microwave detector, then, the larger the width of the resonance, the larger is the range of frequencies for which the detector gives a response to the driving signal. Also, the population inversion can be interpreted as a response to the microwave driving.

\begin{figure}
	\includegraphics[width=\columnwidth]{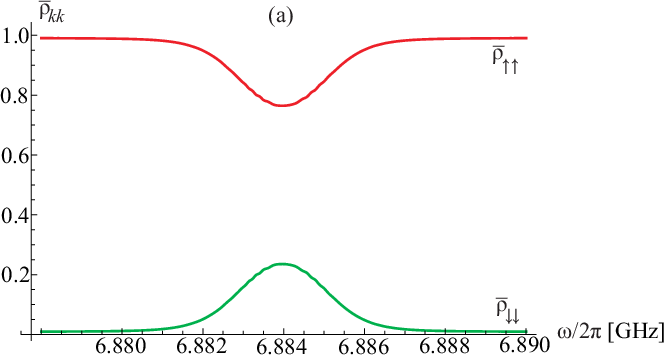}
	
	\vspace{0.2cm}
	\includegraphics[width=\columnwidth]{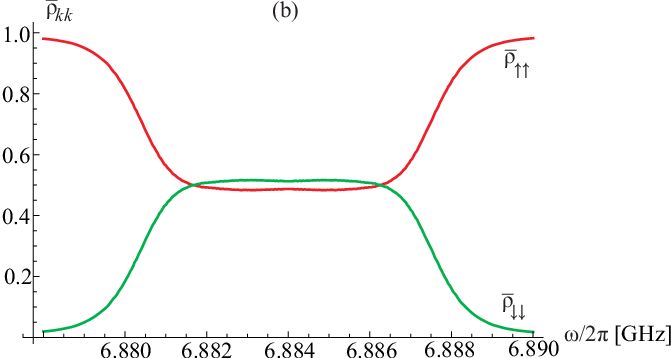}
	
	\vspace{0.2cm}
	\includegraphics[width=\columnwidth]{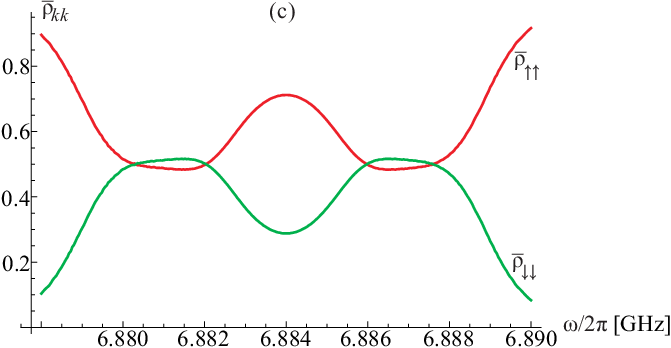}
	
	\vspace{0.2cm}
	\includegraphics[width=\columnwidth]{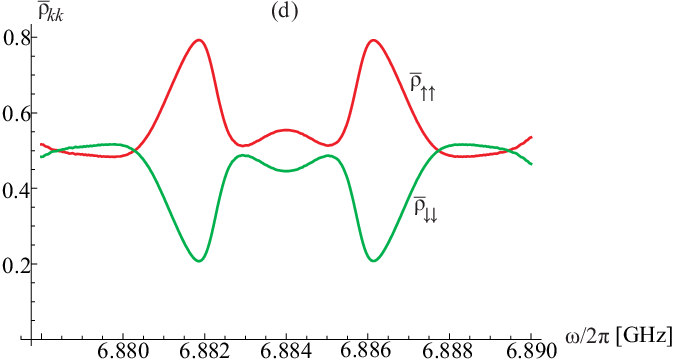}
	\caption{ The time-averaged qubit occupation probabilities as functions of the driving frequency for the case of the frequency-dependent amplitude [as in Eq.~(\ref{A-on-omega})] and for different values of the overall factor $A$. The green and red colors correspond to ${\overline\rho}_{\downarrow\downarrow}$ and ${\overline\rho}_{\uparrow\uparrow}$. The amplitude $A$ is as follows: (a) $30\,\text{GHz}\cdot h$, (b) $48\,\text{GHz}\cdot h$, (c) $60\,\text{GHz}\cdot h$, and (d) $80\,\text{GHz}\cdot h$. Other parameters are the following: the offset $\varepsilon_0=40\,\text{GHz}\cdot h$, the tunneling amplitude $\Delta=7\,\text{GHz}\cdot h$, the relaxation rate $\Gamma=0.004\,\text{GHz}\cdot h$, the resonator frequency $\omega_0/2\pi=6.884\,\text{GHz}$, and the FWHM $\kappa=2\pi\times 0.006\,\text{GHz}$ (the quality factor $Q\approx 1990$).  } \label{Fig:highQ}
\end{figure}

\subsection{Frequency-dependent amplitude}

When a driven qubit is coupled to a resonator, the amplitude received by the whole system has a maximum equal to the driving amplitude $A$ just at the resonator frequency $\omega_0$. For off-resonant driving frequencies the amplitude decreases as a Lorentzian function.
Here, we take into account the frequency dependence of the amplitude and also phase of the driving signal as follows~\cite{oconnell2008}
\begin{equation}\label{2-level-epsilon-nu}
\varepsilon(t)=\varepsilon_0 + A(\omega)\cos[\omega t + \varphi(\omega)],
\end{equation}
where
\begin{equation}\label{A-on-omega}
A(\omega) = \frac{A}{\sqrt{1 + \left[2Q(\omega - \omega_0)/\omega_0\right]^2 }}
\end{equation}
and
\begin{equation}\label{phase-on-omega}
\varphi(\omega)=-\arctan\left[2Q\,\frac{\omega - \omega_0}{\omega_0}\right].
\end{equation}
Here $\omega_0$ is the resonator frequency, and $Q$ is the quality factor, which is related to the full width at half maximum (FWHM, $\kappa$) of the function~(\ref{A-on-omega}) as follows 
\begin{equation}
Q = \sqrt{3}\,\omega_0/\kappa.
\end{equation}
We note that the time-averaged occupation probabilities are identical in the presence of $\varphi(\omega)$ and in the absence of this phase shift. 

Figure~\ref{Fig:highQ} shows the frequency dependences of the time-averaged occupation probabilities for the amplitude given by Eq.~(\ref{A-on-omega}) and different values of the overall factor $A$. The range of frequencies is $(\omega_0 - \kappa,\, \omega_0 + \kappa)$. 

\section{Conclusions}

In summary, we considered the resonant excitation of two coupled qubits and a single qubit system. We numerically solved the GKSL equation and calculated the dynamics of the occupation probabilities of the two-qubit system under influence of a periodic microwave excitation resonant with a pair of levels of this four-level system. If this pair of levels is chosen so that just one qubit changes its state under influence of the excitation, then the dynamics is reduced to the one two-level system. For a single qubit, we compared the numerical results for the time-averaged occupation probabilities in the case of a multiphoton excitation with the analytical results obtained within the rotating-wave approximation and, in particular, commented on the multiphoton Bloch-Siegert shift. We also performed numerical calculations of the time-averaged occupation probabilities in the case where the RWA cannot be applied. 
The multiphoton resonances can be interpreted as a response of the qubit to a microwave signal.
For high driving amplitudes, we obtained the population inversion in the diabatic basis. We also plotted the occupation probabilities for the excitation amplitude dependent on the driving frequency as a Lorentzian function. 
Our theoretical results can be useful for describing microwave detection and for expanding the toolbox of coherent quantum control.

\section*{Acknowledgments}
We would like to thank M.~Grajcar, O.~Yu.~Kitsenko, and A.~I.~Ryzhov for fruitful discussions and help with this work. S.N.S. acknowledges financial support of the National Research Foundation of Ukraine (Grant No.~2025.07/0044). This work was partially supported by NATO Science for Peace and Security Program (Project G5796), the U.S. National Academy of Sciences (NAS) and the Office of Naval Research (ONR) in the framework of the IMPRESS-U project, IEEE program “Magnetism for Ukraine 2025/26” (project number 9918), and the scholarship of the NAS of Ukraine.

\appendix

\section{Rotating-wave approximation for the case of the multiphoton excitation in a two-level system}

We start from the case when the dissipation is absent and the Schr\"odinger equation can be considered. We make the unitary transformation to the rotating coordinate system: $\psi'(t)=U^{-1}(t)\psi(t)$, where $\psi(t)$ and $\psi'(t)$ are the wave functions in the original and rotating frames, respectively, and
\begin{equation}\label{transformation-to-rotating-system}
	U(t)=\exp\left(-\frac{i}{\hbar}\int V(t)dt\right)\equiv \exp\left(i\frac{\eta(t)}{2}\sigma_z\right), 
\end{equation}
with
\begin{equation}
	\eta(t)=\frac{A}{\hbar\omega}\sin(\omega t).
\end{equation}
We find the Hamiltonian in the rotating coordinate system so that to keep the Schr\"odinger equation unchanged: $H'(t)=U^\dag(t)H(t)U(t)-i\hbar U^\dag(t)[dU(t)/dt]$. Straightforward calculations give
\begin{equation}\label{Hamiltonian-rotating-frame}
	H'(t)=U^\dag(t)H_0 U(t)=-\frac{1}{2}\left(\begin{array}{cccc}
		\varepsilon_0 & \Delta\, e^{-i\eta(t)} \\
		\Delta\, e^{i\eta(t)} & -\varepsilon_0 \\
	\end{array}\right).
\end{equation}
Using the Jacobi-Anger expansion
\begin{equation}
	e^{ix\sin\theta}=\sum_{n=-\infty}^{\infty}J_n(x)e^{i n\theta},
\end{equation}
where $J_n(x)$ is the Bessel function of the first kind, the Hamiltonian~(\ref{Hamiltonian-rotating-frame}) can be rewritten as
\begin{align}\label{Hamiltonian-in-rotating-system}
	H'(t)= -\frac{1}{2}\left(\begin{array}{cccc}
		\varepsilon_0 & \sum_{n=-\infty}^{\infty} \Delta_n e^{-i n \omega t} \\
		\sum_{n=-\infty}^{\infty} \Delta_n e^{i n \omega t} & -\varepsilon_0 \\
	\end{array}\right).
\end{align}
Here 
\begin{equation}
\Delta_n=\Delta \cdot J_n(A/\hbar\omega). 
\end{equation}
Applying the rotating-wave approximation means that in Eq.~(\ref{Hamiltonian-in-rotating-system}) just one term in the sums should be retained when solving the Schr\"odinger equation, i.e., the resonant term with $n=-K$:
\begin{equation}\label{Hamiltonian-RWA}
	H_\text{RWA}(t)= -\frac{1}{2}\left(\begin{array}{cccc}
		\varepsilon_0 & (-1)^K\Delta_K e^{i K \omega t} \\
		(-1)^K\Delta_K e^{-i K \omega t} & -\varepsilon_0 \\
	\end{array}\right).
\end{equation}

Taking the dissipation into account phenomenologically, we add to the right-hand side of the Liouville-von Neumann equation for the matrix element $\rho_{kk'}(t)$ the following term: $-(\rho_{kk'}(t)-\rho_{kk'}^{(0)})/\tau_{kk'}$, which gives the exponentially decaying time dependence of $\rho_{kk'}(t)$. The matrix elements of the equilibrium density operator $\rho^{(0)}$ can be evaluated by using the formula
\begin{equation}
\rho^{(0)}=\frac{1}{\Sigma}\exp(-H_0/k_\text{B}T), 
\end{equation}
where $H_0$ is defined in Eq.~(\ref{2-level-H0}) and the statistical sum $\Sigma$ is obtained from the normalization condition $\text{Tr}\rho^{(0)}=1$. As a result, the equilibrium matrix elements of the density operator are the following:
\begin{eqnarray}\label{equilibrium-matrix-elements}
	&\rho_{++}^{(0)}=\frac{1}{\Sigma}\exp\left(-\frac{\delta E}{2k_\text{B}T}\right), \;\; 
	\rho_{--}^{(0)}=\frac{1}{\Sigma}\exp\left(\frac{\delta E}{2k_\text{B}T}\right),\\ 
	&\rho_{+-}^{(0)}=0, 
\end{eqnarray}
where $\Sigma=2\cosh\left(\delta E/2k_\text{B}T\right)$ and $\delta E=\sqrt{\Delta^2+\varepsilon_0^2}$.

Denoting $\tau_{kk}=T_1$ and $\tau_{kk'}=T_2$ $(k\neq k')$, $T_1$ and $T_2$ being the energy relaxation and decoherence times, we write down the differential equations for the diagonal and off-diagonal matrix elements of the density operator as follows
\begin{align}
	&\frac{d\rho_{kk}(t)}{dt}=-\frac{i}{\hbar}\left[H_\text{RWA}(t),\rho(t)\right]_{kk}
	-\frac{\rho_{kk}(t)-\rho_{kk}^{(0)}}{T_1},\\
	&\frac{d\rho_{kk'}(t)}{dt}=-\frac{i}{\hbar}\left[H_\text{RWA}(t),\rho(t)\right]_{kk'}
	-\frac{\rho_{kk'}(t)}{T_2}, \quad k\neq k'.
\end{align}
It is convenient to introduce variables $X$, $Y$, and $Z$, which are defined as follows
\begin{equation}\label{variables-X-Y-Z}
	X+iY=2\rho_{+-}e^{iK\omega t}, \qquad Z=\rho_{--}-\rho_{++}.
\end{equation}
Straightforward calculations lead to the following system of differential equations (the Bloch equations)
\begin{align}\label{system-for-X-Y-Z}
	&\frac{dX}{dt}=\left(-K\omega+\frac{\varepsilon_0}{\hbar}\right)Y-\frac{X}{T_2},
	\nonumber\\
	&\frac{dY}{dt}=\left(K\omega-\frac{\varepsilon_0}{\hbar}\right)X-\frac{Y}{T_2}
	+(-1)^K\frac{\Delta_K}{\hbar}Z,\nonumber\\
	&\frac{dZ}{dt}=(-1)^{K+1}\frac{\Delta_K}{\hbar}Y
	-\frac{Z-Z^{(0)}}{T_1},
\end{align}
where $Z^{(0)}=\rho_{--}^{(0)}-\rho_{++}^{(0)}$. 
Taking into account the expressions for the equilibrium matrix elements~(\ref{equilibrium-matrix-elements}) of the density operator, we obtain in the low-temperature limit $T\to 0$ that $Z^{(0)}=1$. We are interested in the stationary solution of the system~(\ref{system-for-X-Y-Z}), so that the left-hand sides of these equations are equal to zero. This stationary solution reads
\begin{align}
	&X=(-1)^{K+1}\frac{1}{D}(K\hbar\omega-\varepsilon_0)\Delta_K,\label{X-stationary}\\
	&Y=(-1)^K\frac{1}{D}\,\frac{\hbar}{T_2}\Delta_K,\label{Y-stationary}\\
	&Z=1-\frac{1}{D}\,\frac{T_1}{T_2}\Delta_K^2,\label{Z-stationary}
\end{align}
where 
\begin{equation}
	D=(K \hbar\omega-\varepsilon_0)^2 + \left(\frac{\hbar}{T_2}\right)^2+\frac{T_1}{T_2}\Delta_K^2.
\end{equation}

\bibliography{References}

\begin{thebibliography}{44}%
\makeatletter
\providecommand \@ifxundefined [1]{%
 \@ifx{#1\undefined}
}%
\providecommand \@ifnum [1]{%
 \ifnum #1\expandafter \@firstoftwo
 \else \expandafter \@secondoftwo
 \fi
}%
\providecommand \@ifx [1]{%
 \ifx #1\expandafter \@firstoftwo
 \else \expandafter \@secondoftwo
 \fi
}%
\providecommand \natexlab [1]{#1}%
\providecommand \enquote  [1]{``#1''}%
\providecommand \bibnamefont  [1]{#1}%
\providecommand \bibfnamefont [1]{#1}%
\providecommand \citenamefont [1]{#1}%
\providecommand \href@noop [0]{\@secondoftwo}%
\providecommand \href [0]{\begingroup \@sanitize@url \@href}%
\providecommand \@href[1]{\@@startlink{#1}\@@href}%
\providecommand \@@href[1]{\endgroup#1\@@endlink}%
\providecommand \@sanitize@url [0]{\catcode `\\12\catcode `\$12\catcode
  `\&12\catcode `\#12\catcode `\^12\catcode `\_12\catcode `\%12\relax}%
\providecommand \@@startlink[1]{}%
\providecommand \@@endlink[0]{}%
\providecommand \url  [0]{\begingroup\@sanitize@url \@url }%
\providecommand \@url [1]{\endgroup\@href {#1}{\urlprefix }}%
\providecommand \urlprefix  [0]{URL }%
\providecommand \Eprint [0]{\href }%
\providecommand \doibase [0]{https://doi.org/}%
\providecommand \selectlanguage [0]{\@gobble}%
\providecommand \bibinfo  [0]{\@secondoftwo}%
\providecommand \bibfield  [0]{\@secondoftwo}%
\providecommand \translation [1]{[#1]}%
\providecommand \BibitemOpen [0]{}%
\providecommand \bibitemStop [0]{}%
\providecommand \bibitemNoStop [0]{.\EOS\space}%
\providecommand \EOS [0]{\spacefactor3000\relax}%
\providecommand \BibitemShut  [1]{\csname bibitem#1\endcsname}%
\let\auto@bib@innerbib\@empty
\bibitem [{\citenamefont {Silveri}\ \emph {et~al.}(2017)\citenamefont
  {Silveri}, \citenamefont {Tuorila}, \citenamefont {Thuneberg},\ and\
  \citenamefont {Paraoanu}}]{silveri2017}%
  \BibitemOpen
  \bibfield  {author} {\bibinfo {author} {\bibfnamefont {M.~P.}\ \bibnamefont
  {Silveri}}, \bibinfo {author} {\bibfnamefont {J.~A.}\ \bibnamefont
  {Tuorila}}, \bibinfo {author} {\bibfnamefont {E.~V.}\ \bibnamefont
  {Thuneberg}},\ and\ \bibinfo {author} {\bibfnamefont {G.~S.}\ \bibnamefont
  {Paraoanu}},\ }\bibfield  {title} {\bibinfo {title} {Quantum systems under
  frequency modulation},\ }\href@noop {} {\bibfield  {journal} {\bibinfo
  {journal} {{Rep. Prog. Phys.}}\ }\textbf {\bibinfo {volume} {80}},\ \bibinfo
  {pages} {056002} (\bibinfo {year} {2017})}\BibitemShut {NoStop}%
\bibitem [{\citenamefont {Konovalenko}\ and\ \citenamefont
  {Maizelis}(2025)}]{konovalenko2025}%
  \BibitemOpen
  \bibfield  {author} {\bibinfo {author} {\bibfnamefont {O.~M.}\ \bibnamefont
  {Konovalenko}}\ and\ \bibinfo {author} {\bibfnamefont {Z.~A.}\ \bibnamefont
  {Maizelis}},\ }\bibfield  {title} {\bibinfo {title} {Suppression of
  decoherence by multiple joint measurements in entangled systems},\
  }\href@noop {} {\bibfield  {journal} {\bibinfo  {journal} {{Low Temp.
  Phys.}}\ }\textbf {\bibinfo {volume} {51}},\ \bibinfo {pages} {1366}
  (\bibinfo {year} {2025})}\BibitemShut {NoStop}%
\bibitem [{\citenamefont {Izmalkov}\ \emph {et~al.}(2004)\citenamefont
  {Izmalkov}, \citenamefont {Grajcar}, \citenamefont {Il’ichev},
  \citenamefont {{Th. Wagner}}, \citenamefont {Meyer}, \citenamefont {{A. Yu.
  Smirnov}}, \citenamefont {Amin}, \citenamefont {{v}an~{d}en Brink},\ and\
  \citenamefont {Zagoskin}}]{izmalkov2004}%
  \BibitemOpen
  \bibfield  {author} {\bibinfo {author} {\bibfnamefont {A.}~\bibnamefont
  {Izmalkov}}, \bibinfo {author} {\bibfnamefont {M.}~\bibnamefont {Grajcar}},
  \bibinfo {author} {\bibfnamefont {E.}~\bibnamefont {Il’ichev}}, \bibinfo
  {author} {\bibnamefont {{Th. Wagner}}}, \bibinfo {author} {\bibfnamefont
  {H.-G.}\ \bibnamefont {Meyer}}, \bibinfo {author} {\bibnamefont {{A. Yu.
  Smirnov}}}, \bibinfo {author} {\bibfnamefont {M.~H.~S.}\ \bibnamefont
  {Amin}}, \bibinfo {author} {\bibfnamefont {A.~M.}\ \bibnamefont {{v}an~{d}en
  Brink}},\ and\ \bibinfo {author} {\bibfnamefont {A.~M.}\ \bibnamefont
  {Zagoskin}},\ }\bibfield  {title} {\bibinfo {title} {Evidence for entangled
  states of two coupled flux qubits},\ }\href@noop {} {\bibfield  {journal}
  {\bibinfo  {journal} {{Phys. Rev. Lett.}}\ }\textbf {\bibinfo {volume}
  {93}},\ \bibinfo {pages} {037003} (\bibinfo {year} {2004})}\BibitemShut
  {NoStop}%
\bibitem [{\citenamefont {Gramajo}\ \emph
  {et~al.}(2018{\natexlab{a}})\citenamefont {Gramajo}, \citenamefont
  {Dom{\'\i}nguez},\ and\ \citenamefont {S{\'a}nchez}}]{gramajo2018}%
  \BibitemOpen
  \bibfield  {author} {\bibinfo {author} {\bibfnamefont {A.~L.}\ \bibnamefont
  {Gramajo}}, \bibinfo {author} {\bibfnamefont {D.}~\bibnamefont
  {Dom{\'\i}nguez}},\ and\ \bibinfo {author} {\bibfnamefont {M.~J.}\
  \bibnamefont {S{\'a}nchez}},\ }\bibfield  {title} {\bibinfo {title}
  {Amplitude tuning of steady-state entanglement in strongly driven coupled
  qubits},\ }\href@noop {} {\bibfield  {journal} {\bibinfo  {journal} {{Phys.
  Rev. A}}\ }\textbf {\bibinfo {volume} {98}},\ \bibinfo {pages} {042337}
  (\bibinfo {year} {2018}{\natexlab{a}})}\BibitemShut {NoStop}%
\bibitem [{\citenamefont {Antoni{\'c}}\ \emph {et~al.}(2025)\citenamefont
  {Antoni{\'c}}, \citenamefont {Hazanov}, \citenamefont {Masis}, \citenamefont
  {Podolsky},\ and\ \citenamefont {Buks}}]{antonic2025}%
  \BibitemOpen
  \bibfield  {author} {\bibinfo {author} {\bibfnamefont {L.}~\bibnamefont
  {Antoni{\'c}}}, \bibinfo {author} {\bibfnamefont {S.}~\bibnamefont
  {Hazanov}}, \bibinfo {author} {\bibfnamefont {S.}~\bibnamefont {Masis}},
  \bibinfo {author} {\bibfnamefont {D.}~\bibnamefont {Podolsky}},\ and\
  \bibinfo {author} {\bibfnamefont {E.}~\bibnamefont {Buks}},\ }\bibfield
  {title} {\bibinfo {title} {Sideband spectroscopy in the strong driving
  regime: Volcano transparency and sideband anomaly},\ }\href@noop {}
  {\bibfield  {journal} {\bibinfo  {journal} {arXiv:2508.14781}\ } (\bibinfo
  {year} {2025})}\BibitemShut {NoStop}%
\bibitem [{\citenamefont {Besse}\ \emph {et~al.}(2018)\citenamefont {Besse},
  \citenamefont {Gasparinetti}, \citenamefont {Collodo}, \citenamefont
  {Walter}, \citenamefont {Kurpiers}, \citenamefont {Pechal}, \citenamefont
  {Eichler},\ and\ \citenamefont {Wallraff}}]{besse2018}%
  \BibitemOpen
  \bibfield  {author} {\bibinfo {author} {\bibfnamefont {J.-C.}\ \bibnamefont
  {Besse}}, \bibinfo {author} {\bibfnamefont {S.}~\bibnamefont {Gasparinetti}},
  \bibinfo {author} {\bibfnamefont {M.~C.}\ \bibnamefont {Collodo}}, \bibinfo
  {author} {\bibfnamefont {T.}~\bibnamefont {Walter}}, \bibinfo {author}
  {\bibfnamefont {P.}~\bibnamefont {Kurpiers}}, \bibinfo {author}
  {\bibfnamefont {M.}~\bibnamefont {Pechal}}, \bibinfo {author} {\bibfnamefont
  {C.}~\bibnamefont {Eichler}},\ and\ \bibinfo {author} {\bibfnamefont
  {A.}~\bibnamefont {Wallraff}},\ }\bibfield  {title} {\bibinfo {title}
  {Single-shot quantum nondemolition detection of individual itinerant
  microwave photons},\ }\href@noop {} {\bibfield  {journal} {\bibinfo
  {journal} {Phys. Rev. X}\ }\textbf {\bibinfo {volume} {8}},\ \bibinfo {pages}
  {021003} (\bibinfo {year} {2018})}\BibitemShut {NoStop}%
\bibitem [{\citenamefont {Dassonneville}\ \emph {et~al.}(2020)\citenamefont
  {Dassonneville}, \citenamefont {Assouly}, \citenamefont {Peronnin},
  \citenamefont {Rouchon},\ and\ \citenamefont {Huard}}]{dassonneville2020}%
  \BibitemOpen
  \bibfield  {author} {\bibinfo {author} {\bibfnamefont {R.}~\bibnamefont
  {Dassonneville}}, \bibinfo {author} {\bibfnamefont {R.}~\bibnamefont
  {Assouly}}, \bibinfo {author} {\bibfnamefont {T.}~\bibnamefont {Peronnin}},
  \bibinfo {author} {\bibfnamefont {P.}~\bibnamefont {Rouchon}},\ and\ \bibinfo
  {author} {\bibfnamefont {B.}~\bibnamefont {Huard}},\ }\bibfield  {title}
  {\bibinfo {title} {Number-resolved photocounter for propagating microwave
  mode},\ }\href@noop {} {\bibfield  {journal} {\bibinfo  {journal} {Phys. Rev.
  Appl.}\ }\textbf {\bibinfo {volume} {14}},\ \bibinfo {pages} {044022}
  (\bibinfo {year} {2020})}\BibitemShut {NoStop}%
\bibitem [{\citenamefont {Lescanne}\ \emph {et~al.}(2020)\citenamefont
  {Lescanne}, \citenamefont {Del{\'e}glise}, \citenamefont {Albertinale},
  \citenamefont {R{\'e}glade}, \citenamefont {Capelle}, \citenamefont {Ivanov},
  \citenamefont {Jacqmin}, \citenamefont {Leghtas},\ and\ \citenamefont
  {Flurin}}]{lescanne2020}%
  \BibitemOpen
  \bibfield  {author} {\bibinfo {author} {\bibfnamefont {R.}~\bibnamefont
  {Lescanne}}, \bibinfo {author} {\bibfnamefont {S.}~\bibnamefont
  {Del{\'e}glise}}, \bibinfo {author} {\bibfnamefont {E.}~\bibnamefont
  {Albertinale}}, \bibinfo {author} {\bibfnamefont {U.}~\bibnamefont
  {R{\'e}glade}}, \bibinfo {author} {\bibfnamefont {T.}~\bibnamefont
  {Capelle}}, \bibinfo {author} {\bibfnamefont {E.}~\bibnamefont {Ivanov}},
  \bibinfo {author} {\bibfnamefont {T.}~\bibnamefont {Jacqmin}}, \bibinfo
  {author} {\bibfnamefont {Z.}~\bibnamefont {Leghtas}},\ and\ \bibinfo {author}
  {\bibfnamefont {E.}~\bibnamefont {Flurin}},\ }\bibfield  {title} {\bibinfo
  {title} {Irreversible qubit-photon coupling for the detection of itinerant
  microwave photons},\ }\href@noop {} {\bibfield  {journal} {\bibinfo
  {journal} {{Phys. Rev.} X}\ }\textbf {\bibinfo {volume} {10}},\ \bibinfo
  {pages} {021038} (\bibinfo {year} {2020})}\BibitemShut {NoStop}%
\bibitem [{\citenamefont {Opremcak}\ \emph {et~al.}(2018)\citenamefont
  {Opremcak}, \citenamefont {Pechenezhskiy}, \citenamefont {Howington},
  \citenamefont {Christensen}, \citenamefont {Beck}, \citenamefont
  {Leonard~Jr}, \citenamefont {Suttle}, \citenamefont {Wilen}, \citenamefont
  {Nesterov}, \citenamefont {Ribeill} \emph {et~al.}}]{opremcak2018}%
  \BibitemOpen
  \bibfield  {author} {\bibinfo {author} {\bibfnamefont {A.}~\bibnamefont
  {Opremcak}}, \bibinfo {author} {\bibfnamefont {I.~V.}\ \bibnamefont
  {Pechenezhskiy}}, \bibinfo {author} {\bibfnamefont {C.}~\bibnamefont
  {Howington}}, \bibinfo {author} {\bibfnamefont {B.~G.}\ \bibnamefont
  {Christensen}}, \bibinfo {author} {\bibfnamefont {M.~A.}\ \bibnamefont
  {Beck}}, \bibinfo {author} {\bibfnamefont {E.}~\bibnamefont {Leonard~Jr}},
  \bibinfo {author} {\bibfnamefont {J.}~\bibnamefont {Suttle}}, \bibinfo
  {author} {\bibfnamefont {C.}~\bibnamefont {Wilen}}, \bibinfo {author}
  {\bibfnamefont {K.~N.}\ \bibnamefont {Nesterov}}, \bibinfo {author}
  {\bibfnamefont {G.~J.}\ \bibnamefont {Ribeill}}, \emph {et~al.},\ }\bibfield
  {title} {\bibinfo {title} {Measurement of a superconducting qubit with a
  microwave photon counter},\ }\href@noop {} {\bibfield  {journal} {\bibinfo
  {journal} {Science}\ }\textbf {\bibinfo {volume} {361}},\ \bibinfo {pages}
  {1239} (\bibinfo {year} {2018})}\BibitemShut {NoStop}%
\bibitem [{\citenamefont {Ilinskaya}\ \emph {et~al.}(2024)\citenamefont
  {Ilinskaya}, \citenamefont {Ryzhov},\ and\ \citenamefont
  {Shevchenko}}]{ilinskaya2024}%
  \BibitemOpen
  \bibfield  {author} {\bibinfo {author} {\bibfnamefont {O.~A.}\ \bibnamefont
  {Ilinskaya}}, \bibinfo {author} {\bibfnamefont {A.~I.}\ \bibnamefont
  {Ryzhov}},\ and\ \bibinfo {author} {\bibfnamefont {S.~N.}\ \bibnamefont
  {Shevchenko}},\ }\bibfield  {title} {\bibinfo {title} {Flux qubit based
  detector of microwave photons},\ }\href@noop {} {\bibfield  {journal}
  {\bibinfo  {journal} {Phys. Rev. B}\ }\textbf {\bibinfo {volume} {110}},\
  \bibinfo {pages} {155414} (\bibinfo {year} {2024})}\BibitemShut {NoStop}%
\bibitem [{\citenamefont {Stolyarov}\ \emph {et~al.}(2023)\citenamefont
  {Stolyarov}, \citenamefont {Kliushnichenko}, \citenamefont {Kovtoniuk},\ and\
  \citenamefont {Semenov}}]{stolyarov2023}%
  \BibitemOpen
  \bibfield  {author} {\bibinfo {author} {\bibfnamefont {E.~V.}\ \bibnamefont
  {Stolyarov}}, \bibinfo {author} {\bibfnamefont {O.~V.}\ \bibnamefont
  {Kliushnichenko}}, \bibinfo {author} {\bibfnamefont {V.~S.}\ \bibnamefont
  {Kovtoniuk}},\ and\ \bibinfo {author} {\bibfnamefont {A.~A.}\ \bibnamefont
  {Semenov}},\ }\bibfield  {title} {\bibinfo {title} {Photon-number resolution
  with microwave {J}osephson photomultipliers},\ }\href@noop {} {\bibfield
  {journal} {\bibinfo  {journal} {{Phys. Rev.} A}\ }\textbf {\bibinfo {volume}
  {108}},\ \bibinfo {pages} {063710} (\bibinfo {year} {2023})}\BibitemShut
  {NoStop}%
\bibitem [{\citenamefont {Stolyarov}\ and\ \citenamefont
  {Baskov}(2025)}]{stolyarov2025}%
  \BibitemOpen
  \bibfield  {author} {\bibinfo {author} {\bibfnamefont {E.~V.}\ \bibnamefont
  {Stolyarov}}\ and\ \bibinfo {author} {\bibfnamefont {R.~A.}\ \bibnamefont
  {Baskov}},\ }\bibfield  {title} {\bibinfo {title} {Detector of microwave
  photon pairs based on a {J}osephson photomultiplier},\ }\href@noop {}
  {\bibfield  {journal} {\bibinfo  {journal} {{Phys. Rev. Res.}}\ }\textbf
  {\bibinfo {volume} {7}},\ \bibinfo {pages} {033263} (\bibinfo {year}
  {2025})}\BibitemShut {NoStop}%
\bibitem [{\citenamefont {Sirenko}\ and\ \citenamefont
  {Bartolom{\'e}}(2025)}]{sirenko2025}%
  \BibitemOpen
  \bibfield  {author} {\bibinfo {author} {\bibfnamefont {V.}~\bibnamefont
  {Sirenko}}\ and\ \bibinfo {author} {\bibfnamefont {J.}~\bibnamefont
  {Bartolom{\'e}}},\ }\bibfield  {title} {\bibinfo {title} {Molecular spin
  relaxation of f-block metal complexes for quantum applications in a
  nutshell},\ }\href@noop {} {\bibfield  {journal} {\bibinfo  {journal} {Low
  Temp. Phys.}\ }\textbf {\bibinfo {volume} {51}},\ \bibinfo {pages} {517}
  (\bibinfo {year} {2025})}\BibitemShut {NoStop}%
\bibitem [{\citenamefont {Savasta}\ \emph {et~al.}(2021)\citenamefont
  {Savasta}, \citenamefont {Di~Stefano}, \citenamefont {Settineri},
  \citenamefont {Zueco}, \citenamefont {Hughes},\ and\ \citenamefont
  {Nori}}]{savasta2021}%
  \BibitemOpen
  \bibfield  {author} {\bibinfo {author} {\bibfnamefont {S.}~\bibnamefont
  {Savasta}}, \bibinfo {author} {\bibfnamefont {O.}~\bibnamefont {Di~Stefano}},
  \bibinfo {author} {\bibfnamefont {A.}~\bibnamefont {Settineri}}, \bibinfo
  {author} {\bibfnamefont {D.}~\bibnamefont {Zueco}}, \bibinfo {author}
  {\bibfnamefont {S.}~\bibnamefont {Hughes}},\ and\ \bibinfo {author}
  {\bibfnamefont {F.}~\bibnamefont {Nori}},\ }\bibfield  {title} {\bibinfo
  {title} {Gauge principle and gauge invariance in two-level systems},\
  }\href@noop {} {\bibfield  {journal} {\bibinfo  {journal} {{Phys. Rev.} A}\
  }\textbf {\bibinfo {volume} {103}},\ \bibinfo {pages} {053703} (\bibinfo
  {year} {2021})}\BibitemShut {NoStop}%
\bibitem [{\citenamefont {Rettaroli}\ \emph {et~al.}(2025)\citenamefont
  {Rettaroli}, \citenamefont {Banchi}, \citenamefont {Corti}, \citenamefont
  {D’Elia}, \citenamefont {Gatti}, \citenamefont {Giachero}, \citenamefont
  {Labranca}, \citenamefont {Moretti}, \citenamefont {Nucciotti}, \citenamefont
  {Komnang} \emph {et~al.}}]{rettaroli2025}%
  \BibitemOpen
  \bibfield  {author} {\bibinfo {author} {\bibfnamefont {A.}~\bibnamefont
  {Rettaroli}}, \bibinfo {author} {\bibfnamefont {L.}~\bibnamefont {Banchi}},
  \bibinfo {author} {\bibfnamefont {H.~A.}\ \bibnamefont {Corti}}, \bibinfo
  {author} {\bibfnamefont {A.}~\bibnamefont {D’Elia}}, \bibinfo {author}
  {\bibfnamefont {C.}~\bibnamefont {Gatti}}, \bibinfo {author} {\bibfnamefont
  {A.}~\bibnamefont {Giachero}}, \bibinfo {author} {\bibfnamefont
  {D.}~\bibnamefont {Labranca}}, \bibinfo {author} {\bibfnamefont
  {R.}~\bibnamefont {Moretti}}, \bibinfo {author} {\bibfnamefont
  {A.}~\bibnamefont {Nucciotti}}, \bibinfo {author} {\bibfnamefont {A.~S.~P.}\
  \bibnamefont {Komnang}}, \emph {et~al.},\ }\bibfield  {title} {\bibinfo
  {title} {Novel two-qubit microwave photon detector for fundamental physics
  applications},\ }\href@noop {} {\bibfield  {journal} {\bibinfo  {journal}
  {Nucl. Instrum. Methods Phys. Res. A}\ }\textbf {\bibinfo {volume} {1070}},\
  \bibinfo {pages} {170010} (\bibinfo {year} {2025})}\BibitemShut {NoStop}%
\bibitem [{\citenamefont {Reh{\'a}k}\ \emph {et~al.}(2014)\citenamefont
  {Reh{\'a}k}, \citenamefont {Neilinger}, \citenamefont {{\v{Z}}emli{\v{c}}ka},
  \citenamefont {Manca}, \citenamefont {H{\"u}bner}, \citenamefont {Il'ichev},\
  and\ \citenamefont {Grajcar}}]{rehak2014}%
  \BibitemOpen
  \bibfield  {author} {\bibinfo {author} {\bibfnamefont {M.}~\bibnamefont
  {Reh{\'a}k}}, \bibinfo {author} {\bibfnamefont {P.}~\bibnamefont
  {Neilinger}}, \bibinfo {author} {\bibfnamefont {M.}~\bibnamefont
  {{\v{Z}}emli{\v{c}}ka}}, \bibinfo {author} {\bibfnamefont {D.}~\bibnamefont
  {Manca}}, \bibinfo {author} {\bibfnamefont {U.}~\bibnamefont {H{\"u}bner}},
  \bibinfo {author} {\bibfnamefont {E.}~\bibnamefont {Il'ichev}},\ and\
  \bibinfo {author} {\bibfnamefont {M.}~\bibnamefont {Grajcar}},\ }\bibfield
  {title} {\bibinfo {title} {Switching effect in {SQUIDs} coupled by
  {J}osephson junction},\ }in\ \href@noop {} {\emph {\bibinfo {booktitle}
  {Proceedings 20. International Conference on Applied Physics of Condensed
  Matter}}},\ \bibinfo {series and number} {\bibinfo {number}
  {INIS-SK--2017-034}}\ (\bibinfo {year} {2014})\ pp.\ \bibinfo {pages}
  {304--307}\BibitemShut {NoStop}%
\bibitem [{\citenamefont {Neilinger}\ \emph {et~al.}()\citenamefont
  {Neilinger}, \citenamefont {Ilinskaya}, \citenamefont {Rehak}, \citenamefont
  {Baranek}, \citenamefont {Kern}, \citenamefont {Rizvanov}, \citenamefont
  {Turutanov}, \citenamefont {Oelsner}, \citenamefont {Shevchenko},
  \citenamefont {Il’ichev},\ and\ \citenamefont {Grajcar}}]{neilinger2026}%
  \BibitemOpen
  \bibfield  {author} {\bibinfo {author} {\bibfnamefont {P.}~\bibnamefont
  {Neilinger}}, \bibinfo {author} {\bibfnamefont {O.~A.}\ \bibnamefont
  {Ilinskaya}}, \bibinfo {author} {\bibfnamefont {M.}~\bibnamefont {Rehak}},
  \bibinfo {author} {\bibfnamefont {M.}~\bibnamefont {Baranek}}, \bibinfo
  {author} {\bibfnamefont {S.}~\bibnamefont {Kern}}, \bibinfo {author}
  {\bibfnamefont {E.}~\bibnamefont {Rizvanov}}, \bibinfo {author}
  {\bibfnamefont {O.~G.}\ \bibnamefont {Turutanov}}, \bibinfo {author}
  {\bibfnamefont {G.}~\bibnamefont {Oelsner}}, \bibinfo {author} {\bibfnamefont
  {S.~N.}\ \bibnamefont {Shevchenko}}, \bibinfo {author} {\bibfnamefont
  {E.}~\bibnamefont {Il’ichev}},\ and\ \bibinfo {author} {\bibfnamefont
  {M.}~\bibnamefont {Grajcar}},\ }\bibfield  {title} {\bibinfo {title} {Double
  {SQUID} photon counter --- proof of principle},\ }\href@noop {} {\bibinfo
  {journal} {in preparation}\ }\BibitemShut {NoStop}%
\bibitem [{\citenamefont {Shevchenko}\ \emph
  {et~al.}(2008{\natexlab{a}})\citenamefont {Shevchenko}, \citenamefont
  {{v}an~{d}er Ploeg}, \citenamefont {Grajcar}, \citenamefont {Il’ichev},
  \citenamefont {Omelyanchouk},\ and\ \citenamefont {Meyer}}]{shevchenko2008}%
  \BibitemOpen
\bibfield  {journal} {  }\bibfield  {author} {\bibinfo {author} {\bibfnamefont
  {S.~N.}\ \bibnamefont {Shevchenko}}, \bibinfo {author} {\bibfnamefont
  {S.~H.~W.}\ \bibnamefont {{v}an~{d}er Ploeg}}, \bibinfo {author}
  {\bibfnamefont {M.}~\bibnamefont {Grajcar}}, \bibinfo {author} {\bibfnamefont
  {E.}~\bibnamefont {Il’ichev}}, \bibinfo {author} {\bibfnamefont {A.~N.}\
  \bibnamefont {Omelyanchouk}},\ and\ \bibinfo {author} {\bibfnamefont {H.-G.}\
  \bibnamefont {Meyer}},\ }\bibfield  {title} {\bibinfo {title} {Resonant
  excitations of single and two-qubit systems coupled to a tank circuit},\
  }\href@noop {} {\bibfield  {journal} {\bibinfo  {journal} {{Phys. Rev. B}}\
  }\textbf {\bibinfo {volume} {78}},\ \bibinfo {pages} {174527} (\bibinfo
  {year} {2008}{\natexlab{a}})}\BibitemShut {NoStop}%
\bibitem [{\citenamefont {Berns}\ \emph {et~al.}(2006)\citenamefont {Berns},
  \citenamefont {Oliver}, \citenamefont {Valenzuela}, \citenamefont {Shytov},
  \citenamefont {Berggren}, \citenamefont {Levitov},\ and\ \citenamefont
  {Orlando}}]{berns2006}%
  \BibitemOpen
  \bibfield  {author} {\bibinfo {author} {\bibfnamefont {D.~M.}\ \bibnamefont
  {Berns}}, \bibinfo {author} {\bibfnamefont {W.~D.}\ \bibnamefont {Oliver}},
  \bibinfo {author} {\bibfnamefont {S.~O.}\ \bibnamefont {Valenzuela}},
  \bibinfo {author} {\bibfnamefont {A.~V.}\ \bibnamefont {Shytov}}, \bibinfo
  {author} {\bibfnamefont {K.~K.}\ \bibnamefont {Berggren}}, \bibinfo {author}
  {\bibfnamefont {L.~S.}\ \bibnamefont {Levitov}},\ and\ \bibinfo {author}
  {\bibfnamefont {T.~P.}\ \bibnamefont {Orlando}},\ }\bibfield  {title}
  {\bibinfo {title} {Coherent quasiclassical dynamics of a persistent current
  qubit},\ }\href@noop {} {\bibfield  {journal} {\bibinfo  {journal} {{Phys.
  Rev. Lett.}}\ }\textbf {\bibinfo {volume} {97}},\ \bibinfo {pages} {150502}
  (\bibinfo {year} {2006})}\BibitemShut {NoStop}%
\bibitem [{\citenamefont {Huang}\ \emph {et~al.}(2025)\citenamefont {Huang},
  \citenamefont {Luneau}, \citenamefont {Schirk}, \citenamefont {Wallner},
  \citenamefont {Schneider}, \citenamefont {Filipp}, \citenamefont {Liegener},\
  and\ \citenamefont {Rabl}}]{huang2025}%
  \BibitemOpen
  \bibfield  {author} {\bibinfo {author} {\bibfnamefont {L.}~\bibnamefont
  {Huang}}, \bibinfo {author} {\bibfnamefont {J.}~\bibnamefont {Luneau}},
  \bibinfo {author} {\bibfnamefont {J.}~\bibnamefont {Schirk}}, \bibinfo
  {author} {\bibfnamefont {F.}~\bibnamefont {Wallner}}, \bibinfo {author}
  {\bibfnamefont {C.~M.~F.}\ \bibnamefont {Schneider}}, \bibinfo {author}
  {\bibfnamefont {S.}~\bibnamefont {Filipp}}, \bibinfo {author} {\bibfnamefont
  {K.}~\bibnamefont {Liegener}},\ and\ \bibinfo {author} {\bibfnamefont
  {P.}~\bibnamefont {Rabl}},\ }\bibfield  {title} {\bibinfo {title} {Theory of
  multi-photon processes for applications in quantum control},\ }\href@noop {}
  {\bibfield  {journal} {\bibinfo  {journal} {{arXiv:2509.16074}}\ } (\bibinfo
  {year} {2025})}\BibitemShut {NoStop}%
\bibitem [{\citenamefont {Kohler}\ and\ \citenamefont
  {Casado-Pascual}(2026)}]{kohler2026}%
  \BibitemOpen
  \bibfield  {author} {\bibinfo {author} {\bibfnamefont {S.}~\bibnamefont
  {Kohler}}\ and\ \bibinfo {author} {\bibfnamefont {J.}~\bibnamefont
  {Casado-Pascual}},\ }\bibfield  {title} {\bibinfo {title} {Hidden
  time-nonlocal {F}loquet symmetries},\ }\href@noop {} {\bibfield  {journal}
  {\bibinfo  {journal} {arXiv:2601.05783}\ } (\bibinfo {year}
  {2026})}\BibitemShut {NoStop}%
\bibitem [{\citenamefont {Gramajo}\ \emph
  {et~al.}(2018{\natexlab{b}})\citenamefont {Gramajo}, \citenamefont
  {Dom{\'\i}nguez},\ and\ \citenamefont {S{\'a}nchez}}]{gramajo2018_2}%
  \BibitemOpen
  \bibfield  {author} {\bibinfo {author} {\bibfnamefont {A.~L.}\ \bibnamefont
  {Gramajo}}, \bibinfo {author} {\bibfnamefont {D.}~\bibnamefont
  {Dom{\'\i}nguez}},\ and\ \bibinfo {author} {\bibfnamefont {M.~J.}\
  \bibnamefont {S{\'a}nchez}},\ }\bibfield  {title} {\bibinfo {title}
  {Controlling entanglement in the interferometry of driven coupled flux
  qubits},\ }in\ \href@noop {} {\emph {\bibinfo {booktitle} {Journal of
  Physics: Conference Series}}},\ Vol.\ \bibinfo {volume} {969}\ (\bibinfo
  {organization} {IOP Publishing},\ \bibinfo {year} {2018})\ p.\ \bibinfo
  {pages} {012135}\BibitemShut {NoStop}%
\bibitem [{\citenamefont {Wang}\ \emph {et~al.}(2025)\citenamefont {Wang},
  \citenamefont {D'Anjou}, \citenamefont {Gigon}, \citenamefont {Blais},\ and\
  \citenamefont {Blok}}]{wang2025}%
  \BibitemOpen
  \bibfield  {author} {\bibinfo {author} {\bibfnamefont {Z.}~\bibnamefont
  {Wang}}, \bibinfo {author} {\bibfnamefont {B.}~\bibnamefont {D'Anjou}},
  \bibinfo {author} {\bibfnamefont {P.}~\bibnamefont {Gigon}}, \bibinfo
  {author} {\bibfnamefont {A.}~\bibnamefont {Blais}},\ and\ \bibinfo {author}
  {\bibfnamefont {M.~S.}\ \bibnamefont {Blok}},\ }\bibfield  {title} {\bibinfo
  {title} {Probing excited-state dynamics of transmon ionization},\ }\href@noop
  {} {\bibfield  {journal} {\bibinfo  {journal} {arXiv:2505.00639}\ } (\bibinfo
  {year} {2025})}\BibitemShut {NoStop}%
\bibitem [{\citenamefont {F{\'e}chant}\ \emph {et~al.}(2025)\citenamefont
  {F{\'e}chant}, \citenamefont {Dumas}, \citenamefont {B{\'e}n{\^a}tre},
  \citenamefont {Gosling}, \citenamefont {Lenhard}, \citenamefont {Spiecker},
  \citenamefont {Geisert}, \citenamefont {Ihssen}, \citenamefont {Wernsdorfer},
  \citenamefont {D’Anjou} \emph {et~al.}}]{fechant2025}%
  \BibitemOpen
  \bibfield  {author} {\bibinfo {author} {\bibfnamefont {M.}~\bibnamefont
  {F{\'e}chant}}, \bibinfo {author} {\bibfnamefont {M.~F.}\ \bibnamefont
  {Dumas}}, \bibinfo {author} {\bibfnamefont {D.}~\bibnamefont
  {B{\'e}n{\^a}tre}}, \bibinfo {author} {\bibfnamefont {N.}~\bibnamefont
  {Gosling}}, \bibinfo {author} {\bibfnamefont {P.}~\bibnamefont {Lenhard}},
  \bibinfo {author} {\bibfnamefont {M.}~\bibnamefont {Spiecker}}, \bibinfo
  {author} {\bibfnamefont {S.}~\bibnamefont {Geisert}}, \bibinfo {author}
  {\bibfnamefont {S.}~\bibnamefont {Ihssen}}, \bibinfo {author} {\bibfnamefont
  {W.}~\bibnamefont {Wernsdorfer}}, \bibinfo {author} {\bibfnamefont
  {B.}~\bibnamefont {D’Anjou}}, \emph {et~al.},\ }\bibfield  {title}
  {\bibinfo {title} {Offset charge dependence of measurement-induced
  transitions in transmons},\ }\href@noop {} {\bibfield  {journal} {\bibinfo
  {journal} {{Phys. Rev. Lett.}}\ }\textbf {\bibinfo {volume} {135}},\ \bibinfo
  {pages} {180603} (\bibinfo {year} {2025})}\BibitemShut {NoStop}%
\bibitem [{\citenamefont {Neilinger}\ \emph {et~al.}(2016)\citenamefont
  {Neilinger}, \citenamefont {Shevchenko}, \citenamefont {Bog{\'a}r},
  \citenamefont {Reh{\'a}k}, \citenamefont {Oelsner}, \citenamefont {Karpov},
  \citenamefont {H{\"u}bner}, \citenamefont {Astafiev}, \citenamefont
  {Grajcar},\ and\ \citenamefont {Il'ichev}}]{neilinger2016}%
  \BibitemOpen
  \bibfield  {author} {\bibinfo {author} {\bibfnamefont {P.}~\bibnamefont
  {Neilinger}}, \bibinfo {author} {\bibfnamefont {S.~N.}\ \bibnamefont
  {Shevchenko}}, \bibinfo {author} {\bibfnamefont {J.}~\bibnamefont
  {Bog{\'a}r}}, \bibinfo {author} {\bibfnamefont {M.}~\bibnamefont
  {Reh{\'a}k}}, \bibinfo {author} {\bibfnamefont {G.}~\bibnamefont {Oelsner}},
  \bibinfo {author} {\bibfnamefont {D.~S.}\ \bibnamefont {Karpov}}, \bibinfo
  {author} {\bibfnamefont {U.}~\bibnamefont {H{\"u}bner}}, \bibinfo {author}
  {\bibfnamefont {O.}~\bibnamefont {Astafiev}}, \bibinfo {author}
  {\bibfnamefont {M.}~\bibnamefont {Grajcar}},\ and\ \bibinfo {author}
  {\bibfnamefont {E.}~\bibnamefont {Il'ichev}},\ }\bibfield  {title} {\bibinfo
  {title} {Landau-{Z}ener-{S}t{\"u}ckelberg-{M}ajorana lasing in circuit
  quantum electrodynamics},\ }\href@noop {} {\bibfield  {journal} {\bibinfo
  {journal} {Phys. Rev. B}\ }\textbf {\bibinfo {volume} {94}},\ \bibinfo
  {pages} {094519} (\bibinfo {year} {2016})}\BibitemShut {NoStop}%
\bibitem [{\citenamefont {Shevchenko}\ \emph {et~al.}(2014)\citenamefont
  {Shevchenko}, \citenamefont {Oelsner}, \citenamefont {{Ya. S. Greenberg}},
  \citenamefont {Macha}, \citenamefont {Karpov}, \citenamefont {Grajcar},
  \citenamefont {H{\"u}bner}, \citenamefont {Omelyanchouk},\ and\ \citenamefont
  {Il'{i}chev}}]{shevchenko2014}%
  \BibitemOpen
  \bibfield  {author} {\bibinfo {author} {\bibfnamefont {S.~N.}\ \bibnamefont
  {Shevchenko}}, \bibinfo {author} {\bibfnamefont {G.}~\bibnamefont {Oelsner}},
  \bibinfo {author} {\bibnamefont {{Ya. S. Greenberg}}}, \bibinfo {author}
  {\bibfnamefont {P.}~\bibnamefont {Macha}}, \bibinfo {author} {\bibfnamefont
  {D.~S.}\ \bibnamefont {Karpov}}, \bibinfo {author} {\bibfnamefont
  {M.}~\bibnamefont {Grajcar}}, \bibinfo {author} {\bibfnamefont
  {U.}~\bibnamefont {H{\"u}bner}}, \bibinfo {author} {\bibfnamefont {A.~N.}\
  \bibnamefont {Omelyanchouk}},\ and\ \bibinfo {author} {\bibfnamefont
  {E.}~\bibnamefont {Il'{i}chev}},\ }\bibfield  {title} {\bibinfo {title}
  {Amplification and attenuation of a probe signal by doubly dressed states},\
  }\href@noop {} {\bibfield  {journal} {\bibinfo  {journal} {{Phys. Rev. B}}\
  }\textbf {\bibinfo {volume} {89}},\ \bibinfo {pages} {184504} (\bibinfo
  {year} {2014})}\BibitemShut {NoStop}%
\bibitem [{\citenamefont {Shevchenko}\ \emph
  {et~al.}(2008{\natexlab{b}})\citenamefont {Shevchenko}, \citenamefont
  {Omelyanchouk}, \citenamefont {Zagoskin}, \citenamefont {Savel'{e}v},\ and\
  \citenamefont {Nori}}]{shevchenko2008_2}%
  \BibitemOpen
  \bibfield  {author} {\bibinfo {author} {\bibfnamefont {S.~N.}\ \bibnamefont
  {Shevchenko}}, \bibinfo {author} {\bibfnamefont {A.~N.}\ \bibnamefont
  {Omelyanchouk}}, \bibinfo {author} {\bibfnamefont {A.~M.}\ \bibnamefont
  {Zagoskin}}, \bibinfo {author} {\bibfnamefont {S.}~\bibnamefont
  {Savel'{e}v}},\ and\ \bibinfo {author} {\bibfnamefont {F.}~\bibnamefont
  {Nori}},\ }\bibfield  {title} {\bibinfo {title} {Distinguishing quantum from
  classical oscillations in a driven phase qubit},\ }\href@noop {} {\bibfield
  {journal} {\bibinfo  {journal} {{New J. Phys.}}\ }\textbf {\bibinfo {volume}
  {10}},\ \bibinfo {pages} {073026} (\bibinfo {year}
  {2008}{\natexlab{b}})}\BibitemShut {NoStop}%
\bibitem [{\citenamefont {O’Connell}\ \emph {et~al.}(2008)\citenamefont
  {O’Connell}, \citenamefont {Ansmann}, \citenamefont {Bialczak},
  \citenamefont {Hofheinz}, \citenamefont {Katz}, \citenamefont {Lucero},
  \citenamefont {McKenney}, \citenamefont {Neeley}, \citenamefont {Wang},
  \citenamefont {Weig}, \citenamefont {Cleland},\ and\ \citenamefont
  {Martinis}}]{oconnell2008}%
  \BibitemOpen
  \bibfield  {author} {\bibinfo {author} {\bibfnamefont {A.~D.}\ \bibnamefont
  {O’Connell}}, \bibinfo {author} {\bibfnamefont {M.}~\bibnamefont
  {Ansmann}}, \bibinfo {author} {\bibfnamefont {R.~C.}\ \bibnamefont
  {Bialczak}}, \bibinfo {author} {\bibfnamefont {M.}~\bibnamefont {Hofheinz}},
  \bibinfo {author} {\bibfnamefont {N.}~\bibnamefont {Katz}}, \bibinfo {author}
  {\bibfnamefont {E.}~\bibnamefont {Lucero}}, \bibinfo {author} {\bibfnamefont
  {C.}~\bibnamefont {McKenney}}, \bibinfo {author} {\bibfnamefont
  {M.}~\bibnamefont {Neeley}}, \bibinfo {author} {\bibfnamefont
  {H.}~\bibnamefont {Wang}}, \bibinfo {author} {\bibfnamefont {E.~M.}\
  \bibnamefont {Weig}}, \bibinfo {author} {\bibfnamefont {A.~N.}\ \bibnamefont
  {Cleland}},\ and\ \bibinfo {author} {\bibfnamefont {J.~M.}\ \bibnamefont
  {Martinis}},\ }\bibfield  {title} {\bibinfo {title} {Microwave dielectric
  loss at single photon energies and millikelvin temperatures},\ }\href@noop {}
  {\bibfield  {journal} {\bibinfo  {journal} {{Appl. Phys. Lett.}}\ }\textbf
  {\bibinfo {volume} {92}},\ \bibinfo {pages} {112903} (\bibinfo {year}
  {2008})}\BibitemShut {NoStop}%
\bibitem [{\citenamefont {{v}an~der Ploeg}\ \emph {et~al.}(2007)\citenamefont
  {{v}an~der Ploeg}, \citenamefont {Izmalkov}, \citenamefont {van~den Brink},
  \citenamefont {H{\"u}bner}, \citenamefont {Grajcar}, \citenamefont
  {Il’{i}chev}, \citenamefont {Meyer},\ and\ \citenamefont
  {Zagoskin}}]{ploeg2007}%
  \BibitemOpen
  \bibfield  {author} {\bibinfo {author} {\bibfnamefont {S.~H.~W.}\
  \bibnamefont {{v}an~der Ploeg}}, \bibinfo {author} {\bibfnamefont
  {A.}~\bibnamefont {Izmalkov}}, \bibinfo {author} {\bibfnamefont {A.~M.}\
  \bibnamefont {van~den Brink}}, \bibinfo {author} {\bibfnamefont
  {U.}~\bibnamefont {H{\"u}bner}}, \bibinfo {author} {\bibfnamefont
  {M.}~\bibnamefont {Grajcar}}, \bibinfo {author} {\bibfnamefont
  {E.}~\bibnamefont {Il’{i}chev}}, \bibinfo {author} {\bibfnamefont {H.-G.}\
  \bibnamefont {Meyer}},\ and\ \bibinfo {author} {\bibfnamefont {A.~M.}\
  \bibnamefont {Zagoskin}},\ }\bibfield  {title} {\bibinfo {title}
  {Controllable coupling of superconducting flux qubits},\ }\href@noop {}
  {\bibfield  {journal} {\bibinfo  {journal} {Phys. Rev. Lett.}\ }\textbf
  {\bibinfo {volume} {98}},\ \bibinfo {pages} {057004} (\bibinfo {year}
  {2007})}\BibitemShut {NoStop}%
\bibitem [{\citenamefont {Zvyagin}\ and\ \citenamefont
  {Zvyagina}(2025{\natexlab{a}})}]{zvyagin2025}%
  \BibitemOpen
  \bibfield  {author} {\bibinfo {author} {\bibfnamefont {A.~A.}\ \bibnamefont
  {Zvyagin}}\ and\ \bibinfo {author} {\bibfnamefont {G.~A.}\ \bibnamefont
  {Zvyagina}},\ }\bibfield  {title} {\bibinfo {title} {Features of static and
  dynamic characteristics of a rhombohedric paramagnet},\ }\href@noop {}
  {\bibfield  {journal} {\bibinfo  {journal} {{Low Temp. Phys.}}\ }\textbf
  {\bibinfo {volume} {51}},\ \bibinfo {pages} {353} (\bibinfo {year}
  {2025}{\natexlab{a}})}\BibitemShut {NoStop}%
\bibitem [{\citenamefont {Zvyagin}\ and\ \citenamefont
  {Zvyagina}(2025{\natexlab{b}})}]{zvyagin2025_2}%
  \BibitemOpen
  \bibfield  {author} {\bibinfo {author} {\bibfnamefont {A.~A.}\ \bibnamefont
  {Zvyagin}}\ and\ \bibinfo {author} {\bibfnamefont {G.~A.}\ \bibnamefont
  {Zvyagina}},\ }\bibfield  {title} {\bibinfo {title} {Spin dimer as a model of
  a “quantum ferrimagnet”},\ }\href@noop {} {\bibfield  {journal} {\bibinfo
   {journal} {{Low Temp. Phys.}}\ }\textbf {\bibinfo {volume} {51}},\ \bibinfo
  {pages} {908} (\bibinfo {year} {2025}{\natexlab{b}})}\BibitemShut {NoStop}%
\bibitem [{\citenamefont {Bahrova}\ \emph {et~al.}(2022)\citenamefont
  {Bahrova}, \citenamefont {Kulinich}, \citenamefont {Gorelik}, \citenamefont
  {Shekhter},\ and\ \citenamefont {Park}}]{bahrova2022}%
  \BibitemOpen
  \bibfield  {author} {\bibinfo {author} {\bibfnamefont {O.~M.}\ \bibnamefont
  {Bahrova}}, \bibinfo {author} {\bibfnamefont {S.~I.}\ \bibnamefont
  {Kulinich}}, \bibinfo {author} {\bibfnamefont {L.~Y.}\ \bibnamefont
  {Gorelik}}, \bibinfo {author} {\bibfnamefont {R.~I.}\ \bibnamefont
  {Shekhter}},\ and\ \bibinfo {author} {\bibfnamefont {H.~C.}\ \bibnamefont
  {Park}},\ }\bibfield  {title} {\bibinfo {title} {Cooling of nanomechanical
  vibrations by {A}ndreev injection},\ }\href@noop {} {\bibfield  {journal}
  {\bibinfo  {journal} {{Low Temp. Phys.}}\ }\textbf {\bibinfo {volume} {48}},\
  \bibinfo {pages} {476} (\bibinfo {year} {2022})}\BibitemShut {NoStop}%
\bibitem [{\citenamefont {Schleich}(2015)}]{schleich2015}%
  \BibitemOpen
  \bibfield  {author} {\bibinfo {author} {\bibfnamefont {W.~P.}\ \bibnamefont
  {Schleich}},\ }\href@noop {} {\emph {\bibinfo {title} {Quantum optics in
  phase space}}}\ (\bibinfo  {publisher} {John Wiley \& Sons},\ \bibinfo {year}
  {2015})\BibitemShut {NoStop}%
\bibitem [{\citenamefont {Blum}(2012)}]{blum2012}%
  \BibitemOpen
  \bibfield  {author} {\bibinfo {author} {\bibfnamefont {K.}~\bibnamefont
  {Blum}},\ }\href@noop {} {\emph {\bibinfo {title} {Density matrix theory and
  applications}}},\ Vol.~\bibinfo {volume} {64}\ (\bibinfo  {publisher}
  {Springer Science \& Business Media},\ \bibinfo {year} {2012})\BibitemShut
  {NoStop}%
\bibitem [{\citenamefont {Temchenko}\ \emph {et~al.}(2011)\citenamefont
  {Temchenko}, \citenamefont {Shevchenko},\ and\ \citenamefont
  {Omelyanchouk}}]{temchenko2011}%
  \BibitemOpen
  \bibfield  {author} {\bibinfo {author} {\bibfnamefont {E.~A.}\ \bibnamefont
  {Temchenko}}, \bibinfo {author} {\bibfnamefont {S.~N.}\ \bibnamefont
  {Shevchenko}},\ and\ \bibinfo {author} {\bibfnamefont {A.~N.}\ \bibnamefont
  {Omelyanchouk}},\ }\bibfield  {title} {\bibinfo {title} {Dissipative dynamics
  of a two-qubit system: Four-level lasing},\ }\href@noop {} {\bibfield
  {journal} {\bibinfo  {journal} {Phys. Rev. B}\ }\textbf {\bibinfo {volume}
  {83}},\ \bibinfo {pages} {144507} (\bibinfo {year} {2011})}\BibitemShut
  {NoStop}%
\bibitem [{\citenamefont {De{\c{c}}ordi}\ and\ \citenamefont
  {Vidiella-Barranco}(2017)}]{decordi2017}%
  \BibitemOpen
  \bibfield  {author} {\bibinfo {author} {\bibfnamefont {G.~L.}\ \bibnamefont
  {De{\c{c}}ordi}}\ and\ \bibinfo {author} {\bibfnamefont {A.}~\bibnamefont
  {Vidiella-Barranco}},\ }\bibfield  {title} {\bibinfo {title} {Two coupled
  qubits interacting with a thermal bath: A comparative study of different
  models},\ }\href@noop {} {\bibfield  {journal} {\bibinfo  {journal} {Opt.
  Commun.}\ }\textbf {\bibinfo {volume} {387}},\ \bibinfo {pages} {366}
  (\bibinfo {year} {2017})}\BibitemShut {NoStop}%
\bibitem [{\citenamefont {Vadimov}\ \emph {et~al.}(2021)\citenamefont
  {Vadimov}, \citenamefont {Tuorila}, \citenamefont {Orell}, \citenamefont
  {Stockburger}, \citenamefont {Ala-Nissila}, \citenamefont {Ankerhold},\ and\
  \citenamefont {M{\"o}tt{\"o}nen}}]{vadimov2021}%
  \BibitemOpen
  \bibfield  {author} {\bibinfo {author} {\bibfnamefont {V.}~\bibnamefont
  {Vadimov}}, \bibinfo {author} {\bibfnamefont {J.}~\bibnamefont {Tuorila}},
  \bibinfo {author} {\bibfnamefont {T.}~\bibnamefont {Orell}}, \bibinfo
  {author} {\bibfnamefont {J.}~\bibnamefont {Stockburger}}, \bibinfo {author}
  {\bibfnamefont {T.}~\bibnamefont {Ala-Nissila}}, \bibinfo {author}
  {\bibfnamefont {J.}~\bibnamefont {Ankerhold}},\ and\ \bibinfo {author}
  {\bibfnamefont {M.}~\bibnamefont {M{\"o}tt{\"o}nen}},\ }\bibfield  {title}
  {\bibinfo {title} {Validity of {B}orn-{M}arkov master equations for single-
  and two-qubit systems},\ }\href@noop {} {\bibfield  {journal} {\bibinfo
  {journal} {Phys. Rev. B}\ }\textbf {\bibinfo {volume} {103}},\ \bibinfo
  {pages} {214308} (\bibinfo {year} {2021})}\BibitemShut {NoStop}%
\bibitem [{\citenamefont {Garziano}\ \emph {et~al.}(2016)\citenamefont
  {Garziano}, \citenamefont {Macr{\`\i}}, \citenamefont {Stassi}, \citenamefont
  {Di~Stefano}, \citenamefont {Nori},\ and\ \citenamefont
  {Savasta}}]{garziano2016}%
  \BibitemOpen
  \bibfield  {author} {\bibinfo {author} {\bibfnamefont {L.}~\bibnamefont
  {Garziano}}, \bibinfo {author} {\bibfnamefont {V.}~\bibnamefont
  {Macr{\`\i}}}, \bibinfo {author} {\bibfnamefont {R.}~\bibnamefont {Stassi}},
  \bibinfo {author} {\bibfnamefont {O.}~\bibnamefont {Di~Stefano}}, \bibinfo
  {author} {\bibfnamefont {F.}~\bibnamefont {Nori}},\ and\ \bibinfo {author}
  {\bibfnamefont {S.}~\bibnamefont {Savasta}},\ }\bibfield  {title} {\bibinfo
  {title} {One photon can simultaneously excite two or more atoms},\
  }\href@noop {} {\bibfield  {journal} {\bibinfo  {journal} {{Phys. Rev.
  Lett.}}\ }\textbf {\bibinfo {volume} {117}},\ \bibinfo {pages} {043601}
  (\bibinfo {year} {2016})}\BibitemShut {NoStop}%
\bibitem [{\citenamefont {Pietik{\"a}inen}\ \emph {et~al.}(2018)\citenamefont
  {Pietik{\"a}inen}, \citenamefont {Danilin}, \citenamefont {Kumar},
  \citenamefont {Tuorila},\ and\ \citenamefont {Paraoanu}}]{pietikainen2018}%
  \BibitemOpen
  \bibfield  {author} {\bibinfo {author} {\bibfnamefont {I.}~\bibnamefont
  {Pietik{\"a}inen}}, \bibinfo {author} {\bibfnamefont {S.}~\bibnamefont
  {Danilin}}, \bibinfo {author} {\bibfnamefont {K.~S.}\ \bibnamefont {Kumar}},
  \bibinfo {author} {\bibfnamefont {J.}~\bibnamefont {Tuorila}},\ and\ \bibinfo
  {author} {\bibfnamefont {G.~S.}\ \bibnamefont {Paraoanu}},\ }\bibfield
  {title} {\bibinfo {title} {Multilevel effects in a driven generalized {Rabi}
  model},\ }\href@noop {} {\bibfield  {journal} {\bibinfo  {journal} {{J. Low
  Temp. Phys.}}\ }\textbf {\bibinfo {volume} {191}},\ \bibinfo {pages} {354}
  (\bibinfo {year} {2018})}\BibitemShut {NoStop}%
\bibitem [{\citenamefont {Reparaz}\ \emph {et~al.}(2025)\citenamefont
  {Reparaz}, \citenamefont {S{\'a}nchez}, \citenamefont {Gatto}, \citenamefont
  {Dominguez},\ and\ \citenamefont {Tosi}}]{reparaz2025}%
  \BibitemOpen
  \bibfield  {author} {\bibinfo {author} {\bibfnamefont {V.}~\bibnamefont
  {Reparaz}}, \bibinfo {author} {\bibfnamefont {M.~J.}\ \bibnamefont
  {S{\'a}nchez}}, \bibinfo {author} {\bibfnamefont {M.}~\bibnamefont {Gatto}},
  \bibinfo {author} {\bibfnamefont {D.}~\bibnamefont {Dominguez}},\ and\
  \bibinfo {author} {\bibfnamefont {L.}~\bibnamefont {Tosi}},\ }\bibfield
  {title} {\bibinfo {title} {Landau-{Z}ener-{S}t{\"u}ckelberg spectroscopy of a
  fluxonium quantum circuit},\ }\href@noop {} {\bibfield  {journal} {\bibinfo
  {journal} {{Phys. Rev. B}}\ }\textbf {\bibinfo {volume} {112}},\ \bibinfo
  {pages} {054517} (\bibinfo {year} {2025})}\BibitemShut {NoStop}%
\bibitem [{\citenamefont {Likharev}(2019)}]{likharev}%
  \BibitemOpen
  \bibfield  {author} {\bibinfo {author} {\bibfnamefont {K.~K.}\ \bibnamefont
  {Likharev}},\ }\href {https://doi.org/10.1088/2053-2563/aaf3a3} {\emph
  {\bibinfo {title} {Quantum Mechanics: Lecture notes}}},\ 2053-2563\ (\bibinfo
   {publisher} {IOP Publishing},\ \bibinfo {year} {2019})\BibitemShut {NoStop}%
\bibitem [{\citenamefont {Shevchenko}(2019)}]{shevchenko_book}%
  \BibitemOpen
  \bibfield  {author} {\bibinfo {author} {\bibfnamefont {S.~N.}\ \bibnamefont
  {Shevchenko}},\ }\href {https://doi.org/10.1142/11310} {\emph {\bibinfo
  {title} {Mesoscopic Physics Meets Quantum Engineering}}}\ (\bibinfo
  {publisher} {World Scientific},\ \bibinfo {year} {2019})\BibitemShut
  {NoStop}%
\bibitem [{\citenamefont {Oliver}\ \emph {et~al.}(2005)\citenamefont {Oliver},
  \citenamefont {Yu}, \citenamefont {Lee}, \citenamefont {Berggren},
  \citenamefont {Levitov},\ and\ \citenamefont {Orlando}}]{oliver2005}%
  \BibitemOpen
  \bibfield  {author} {\bibinfo {author} {\bibfnamefont {W.~D.}\ \bibnamefont
  {Oliver}}, \bibinfo {author} {\bibfnamefont {Y.}~\bibnamefont {Yu}}, \bibinfo
  {author} {\bibfnamefont {J.~C.}\ \bibnamefont {Lee}}, \bibinfo {author}
  {\bibfnamefont {K.~K.}\ \bibnamefont {Berggren}}, \bibinfo {author}
  {\bibfnamefont {L.~S.}\ \bibnamefont {Levitov}},\ and\ \bibinfo {author}
  {\bibfnamefont {T.~P.}\ \bibnamefont {Orlando}},\ }\bibfield  {title}
  {\bibinfo {title} {Mach-{Z}ehnder interferometry in a strongly driven
  superconducting qubit},\ }\href@noop {} {\bibfield  {journal} {\bibinfo
  {journal} {Science}\ }\textbf {\bibinfo {volume} {310}},\ \bibinfo {pages}
  {1653} (\bibinfo {year} {2005})}\BibitemShut {NoStop}%
\bibitem [{\citenamefont {Ivakhnenko}\ \emph {et~al.}(2023)\citenamefont
  {Ivakhnenko}, \citenamefont {Shevchenko},\ and\ \citenamefont
  {Nori}}]{ivakhnenko2023}%
  \BibitemOpen
  \bibfield  {author} {\bibinfo {author} {\bibfnamefont {O.~V.}\ \bibnamefont
  {Ivakhnenko}}, \bibinfo {author} {\bibfnamefont {S.~N.}\ \bibnamefont
  {Shevchenko}},\ and\ \bibinfo {author} {\bibfnamefont {F.}~\bibnamefont
  {Nori}},\ }\bibfield  {title} {\bibinfo {title} {Nonadiabatic
  {L}andau--{Z}ener--{S}t{\"u}ckelberg--{M}ajorana transitions, dynamics, and
  interference},\ }\href@noop {} {\bibfield  {journal} {\bibinfo  {journal}
  {Phys. Rep.}\ }\textbf {\bibinfo {volume} {995}},\ \bibinfo {pages} {1}
  (\bibinfo {year} {2023})}\BibitemShut {NoStop}%
\end{thebibliography}%

	
	
	
	
	
	
	

\end{document}